\documentclass[sigconf]{acmart}

\usepackage{multirow}
\usepackage{caption}
\usepackage{subcaption}
\usepackage{xcolor}
\usepackage{colortbl}
\usepackage{balance}

\AtBeginDocument{%
  }

\setcopyright{acmlicensed}

\copyrightyear{2026}
\acmYear{2026}
\setcopyright{cc}
\setcctype{by-nc-nd}
\acmConference[CIKM '26]{Proceedings of the 35th ACM International Conference on Information and Knowledge Management}{November 07--11, 2026}{Rome, Italy}
\acmBooktitle{Proceedings of the 35th ACM International Conference on Information and Knowledge Management (CIKM '26), November 07--11, 2026, Rome, Italy}
\acmDOI{10.1145/3799682.3841006}
\acmISBN{979-8-4007-2539-5/2026/11}

\begin{document}

\title{Mitigating Popularity Bias in Sequential Recommendations with Sparse Prior-Regularized Objective}

\author{Anton Pembek}
\orcid{0009-0005-1757-3379}
\affiliation{
  \institution{Sber AI Lab}
  \city{Moscow}
  \country{Russian Federation}
}
\affiliation{
  \institution{Lomonosov Moscow State University (MSU)}
  \city{Moscow}
  \country{Russian Federation}
}
\email{apembek@bk.ru}

\author{Anton Klenitskiy}
\orcid{0009-0005-8961-6921}
\affiliation{
  \institution{Sber AI Lab}
  \city{Moscow}
  \country{Russian Federation}
}
\email{antklen@gmail.com}

\author{Alexander Savchenko}
\orcid{0009-0004-8826-1455}
\affiliation{
  \institution{Lomonosov Moscow State University (MSU)}
  \city{Moscow}
  \country{Russian Federation}
}
\email{a.m.savchenko.new@gmail.com}

\author{Alexey Vasilev}
\orcid{0009-0007-1415-2004}
\affiliation{
  \institution{Sber AI Lab}
  \city{Moscow}
  \country{Russian Federation}
}
\affiliation{
  \institution{HSE University}
  \city{Moscow}
  \country{Russian Federation}
}
\email{alexxl.vasilev@yandex.ru}

\renewcommand{\shortauthors}{Anton Pembek, Anton Klenitskiy, Alexander Savchenko, and Alexey Vasilev}

\begin{abstract}

Popularity bias is a well-known challenge in recommender systems, often leading to the over-exposure of popular items and reduced catalog coverage. In this work, we propose PR-entmax, a sparse prior-regularized objective that makes popularity correction context-dependent. We derive our method from a generalized variational principle in which an item popularity prior is incorporated directly into a sparse probability mapping, yielding a popularity-aware active support: the set of items receiving non-zero probabilities and hence non-zero gradients is determined jointly by contextual relevance and item popularity. As a result, the prior affects not only item probabilities, but also which items participate in gradient updates during training. The method requires no architectural changes and can be integrated into existing training pipelines by modifying only the loss function. At inference time, the prior term is removed, and items are ranked by the learned contextual relevance alone. Experiments across multiple benchmark datasets and sequential recommendation architectures show that PR-entmax consistently reduces popularity bias while preserving or improving recommendation relevance, advancing the relevance-debiasing trade-off.

\end{abstract}



\maketitle

\section{Introduction}

Modern recommender systems are typically optimized to maximize ranking or accuracy metrics such as NDCG and Recall. However, this optimization often leads to popularity bias, the systematic over-exposure of head items and under-representation of the long tail~\cite{klimashevskaia2024survey,jannach2015recommenders}. As a result, recommendations tend to concentrate on already popular content, limiting diversity, reinforcing exposure feedback loops, and providing a suboptimal experience for users with niche interests. The problem is particularly relevant in sequential recommendations, where next-item prediction is performed over large, highly imbalanced catalogs using standard likelihood-based training objectives.

A range of approaches has been proposed to mitigate popularity bias, including re-weighting or re-sampling training data~\cite{cremonesi2014recommending,gupta2024eqbal,steck2011item}, incorporating regularization terms that penalize exposure imbalance~\cite{abdollahpouri2017controlling,zhu2021popularity,boratto2021connecting}, applying logit adjustment based on item priors~\cite{menon2020long,cao2019learning}, and post-processing ranking lists to promote long-tail items~\cite{zhu2021popularity,klimashevskaia2022mitigating}. While these methods can reduce the dominance of head items in recommendations, they often introduce a trade-off between bias reduction and relevance. In practice, suppressing popular items may lead to degraded ranking accuracy, as popularity is often correlated with relevance signal. Consequently, mitigating popularity bias without sacrificing recommendation performance remains a challenging problem.

We propose \textit{PR-entmax}, a method for reducing popularity bias in sequential recommendation models that are typically trained with objectives based on softmax or its sampled variants~\cite{sun2019bert4rec,klenitskiy2023turning,wilm2023scaling,tikhonovich2025esasrec}. Our central idea is to make popularity correction context-dependent. We derive a sparse prior-regularized objective from a generalized variational principle, obtaining a sparse probability mapping whose active support is jointly determined by contextual relevance and item popularity. The sparsity is provided by entmax~\cite{peters2019sparse}, which assigns non-zero probability only to items with sufficiently large logits and effectively acts as a differentiable form of hard negative selection~\cite{mezentsev2024scalable,wilm2023scaling}. By incorporating item prior directly into the variational principle, we make this selection popularity-aware: popular items are more likely to enter the competition, but only when they are contextually plausible. Thus, the prior influences that items receive a gradient signal, and the model is trained to discriminate the target item from a dynamically selected set of negatives that are both contextually relevant and popular.

We evaluate PR-entmax across five datasets and two widely-used sequential recommendation architectures (SASRec~\cite{kang2018self} and GRU4Rec~\cite{hidasi2015session}), demonstrating that it consistently mitigates popularity bias while preserving or improving ranking quality. The proposed approach modifies only the training objective and requires no architectural changes. At inference, the prior term is removed, and items are ranked by the learned contextual relevance alone. These properties make it straightforward to integrate our method into existing sequential recommendation pipelines.

Our contributions can be summarized as follows:
\begin{itemize}
    \item We introduce PR-entmax, a sparse prior-regularized objective that makes popularity correction context-dependent. The proposed objective allows item popularity to affect the active support of the loss, i.e., the set of items receiving non-zero probabilities and gradients during training.
    \item We derive the proposed objective from a generalized variational principle that unifies several existing training objectives as special cases, providing a principled foundation for context-dependent debiasing.
    \item Through experiments across five datasets and two sequential architectures, we show that the method advances the relevance–debiasing trade-off. Additionally, we analyze how the sparsity parameter controls this trade-off, identifying regimes with strong performance.
\end{itemize}

\section{Related Work}

\paragraph{Popularity bias mitigation}
Approaches to mitigating popularity bias are commonly grouped into pre-processing, in-processing, and post-processing methods~\cite{klimashevskaia2024survey}. Pre-processing methods modify the training data distribution through re-weighting, re-sampling, or balancing strategies, reducing the dominance of head items before model training~\cite{cremonesi2014recommending,steck2011item,gupta2024eqbal}. Post-processing methods instead modify the final recommendation lists to improve long-tail exposure, diversity, or calibration without changing the trained model~\cite{zhu2021popularity,klimashevskaia2022mitigating,liakhnovich2025smmr}. In-processing methods, which our work belongs to, modify the training procedure itself: examples include regularization terms that penalize exposure imbalance~\cite{abdollahpouri2017controlling,zhu2021popularity,boratto2021connecting} and corrections based on item priors~\cite{menon2020long,cao2019learning}. While these directions can reduce the dominance of popular items, they often introduce a relevance-debiasing trade-off: aggressive promotion of less popular items may reduce ranking quality when popularity is correlated with user preference.

\paragraph{Debiasing through the training objective}
Most relevant to our work are methods that mitigate popularity bias by modifying the loss function. Inverse propensity weighting~\cite{schnabel2016recommendations} re-weights training examples to counteract exposure imbalance. Logit adjustment~\cite{menon2020long} shifts logits by the log-prior of each item or class, training the model to estimate a de-popularized score. It has been studied in long-tailed classification and imbalanced learning~\cite{menon2020long,cao2019learning}. In recommendation, related corrections also appear in sampled softmax and log-Q correction, where item sampling probabilities are accounted for during training~\cite{yi2019sampling,khrylchenko2025correcting}. Margin-based methods~\cite{cao2019learning,zhang2022incorporating} introduce popularity-dependent margins. Other recent approaches include causal interventions based on popularity~\cite{zhang2021causal} and popularity-aware negative sampling that control which negative items are used during training~\cite{prakash2024evaluating}. These methods typically apply their corrections within a dense softmax, where every item contributes to the loss. Our approach instead operates over a sparse mapping, where the correction interacts with a context-dependent subset of competing items.

\paragraph{Sparse probability mappings}
A separate line of work studies sparse probability mappings as alternatives to softmax. Sparsemax and entmax replace dense softmax distributions with sparse distributions, assigning exactly zero probability to part of the output space~\cite{martins2016softmax,peters2019sparse}. These mappings can be derived from the variational principle and Fenchel--Young loss frameworks~\cite{blondel2019learning,blondel2020learning,martins2020sparse,martins2022sparse}. Entmax and related sparse mappings have been applied as an alternative to softmax in attention mechanisms and output distributions~\cite{peters2019sparse,roulet2025loss,martins2020sparse,martins2022sparse}. In recommender systems, sparse mappings have been used mainly inside model architectures, for example, in sparse attention mechanisms~\cite{yuan2021dual} or for combining multiple intent representations~\cite{choi2024multi}. To our knowledge, however, their use as output probability mappings for training sequential recommenders remains underexplored.

\section{Background}

This section introduces the technical background required for our formulation. We first review softmax cross-entropy, the standard training objective in sequential recommendation, and then describe entmax and the Fenchel–Young framework for sparse probability mappings.

\subsection{Softmax and Cross-entropy}
\label{sec:softmax}

Sequential recommendation models such as SASRec~\cite{kang2018self},  \break BERT4Rec~\cite{sun2019bert4rec}, and GRU4Rec~\cite{hidasi2015session} are typically trained using the cross-entropy (CE) loss, either over the full item catalog or with sampled negative items~\cite{klenitskiy2023turning,tikhonovich2025esasrec}. Given a user interaction sequence, the model produces logits $\mathbf{z} \in \mathbb{R}^{|\mathcal{I}|}$, where $\mathcal{I}$ is a set of items and each logit $z_i$ represents the relevance score of item $i$. The objective is to distinguish the ground-truth next item from all other items in the catalog. Prior work has shown that softmax-based CE training substantially outperforms binary cross-entropy (BCE) loss in sequential recommendation settings~\cite{klenitskiy2023turning,wilm2023scaling,petrov2023gsasrec}, and CE training has become the standard optimization objective for modern sequential recommenders.

To obtain a probability distribution over items, logits are transformed using the softmax function:
\begin{equation}
p_i(\mathbf{z})=\frac{\exp(z_i)}{\sum_{j \in \mathcal{I}} \exp(z_j)}
\label{eq:softmax}
\end{equation}
The softmax mapping produces a valid probability distribution, ensuring that $p_i(\mathbf{z}) > 0$ for all items and that probabilities sum to one. Let $y \in \mathcal{I}$ denote the ground-truth next item, and let $\mathbf{y}$ denote the corresponding one-hot vector with $y_i = 1[i=y]$. The CE loss is defined as the negative log-likelihood of the ground-truth item $y$:
\begin{equation}
\mathcal{L}_{\mathrm{CE}}(\mathbf{z},y) = -\log p_y(\mathbf{z})
\label{eq:ce_loss}
\end{equation}
The gradient of the CE loss with respect to the logits has the well-known form
\begin{equation}
\frac{\partial \mathcal{L}_{\mathrm{CE}}}{\partial z_i} =
p_i(\mathbf{z})- y_i,
\label{eq:ce_gradient}
\end{equation}
where $y_i$ is equal to one for the target item and zero otherwise. For the positive target item, the gradient equals $p_y - 1$, which increases the corresponding logit during optimization. For every negative item $i \neq y$, the gradient equals $p_i$, pushing its score downward proportionally to its predicted probability. Consequently, optimization reallocates probability mass from negative items toward the ground-truth item.

An important property of softmax is its full support: every item receives strictly positive probability regardless of its relevance to the current context. This induces dense competition over the entire item catalog. Every negative item participates in the loss computation and receives a non-zero gradient update on every training example, so gradient mass is spread across a very large number of irrelevant negatives. This observation motivates sparse probability mappings, which replace dense global competition with context-dependent sparse competition among a subset of candidate items.

\subsection{Sparse Probability Mapping}
\label{sec:entmax}

\begin{figure*}[t]
    \centering
    \includegraphics[width=0.8\textwidth]{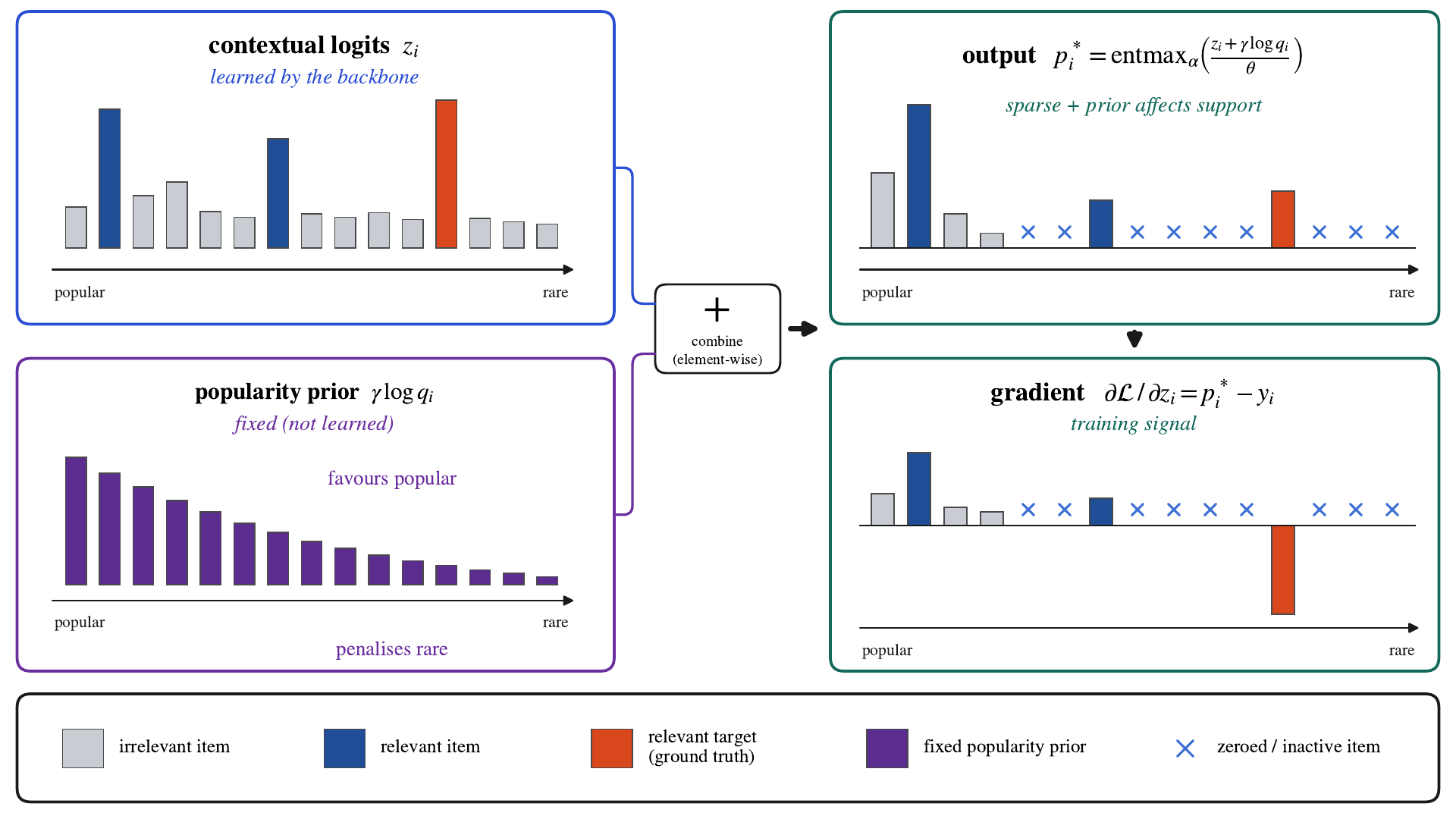}
    \caption{Illustration of the PR-entmax training pipeline. For each item, the contextual logit produced by the backbone model is combined with the corresponding popularity prior. The resulting scores are transformed by entmax into a sparse, popularity-aware output distribution. The active support of this distribution determines which items receive non-zero gradients during training.}
    \label{fig:illustration}
\end{figure*}

Recently, \textit{entmax} was introduced as a sparse generalization of softmax~\cite{peters2019sparse}. This is a family of transformations from logits $\mathbf{z}$ to probabilities $\mathbf{p}$ that produces sparse distributions with exactly zero probabilities for part of the outputs. Entmax is derived from the following variational principle:
\begin{equation}
\label{eq:entmax_mapping}
    p^*(\mathbf{z}) = \text{entmax}_\alpha(\mathbf{z})
    =
    \arg\max_{\mathbf{p} \in \Delta}
    \left\{
    \langle \mathbf{p}, \mathbf{z} \rangle + \mathcal{H}_\alpha(\mathbf{p})
    \right\}.
\end{equation}
where
$\Delta
=
\left\{
\mathbf{p} \in \mathbb{R}^{|\mathcal{I}|}:
p_i \ge 0 \ ,
\sum_{i \in \mathcal{I}} p_i = 1
\right\}$
is the probability simplex,
$\langle \mathbf{p}, \mathbf{z} \rangle=\sum_{i \in \mathcal{I}} p_i z_i$ denotes the dot product,
and $\mathcal{H}_\alpha(\mathbf{p})$ is the Tsallis entropy~\cite{tsallis1988possible}, a generalization of classic Shannon entropy:
\begin{equation}
    \mathcal{H}_\alpha(\mathbf{p})
=
\frac{1}{\alpha(\alpha - 1)}
\left(
1 - \sum_i p_i^\alpha
\right).
\end{equation}
Tsallis entropy is a family of entropies parametrized by $\alpha > 1$. It converges to Shannon entropy $\mathcal{H}(\mathbf{p}) = - \sum_i p_i \log p_i$ in the limit $\alpha \to 1$, and equals Gini entropy for $\alpha = 2$.

In the variational principle, entropy acts as a regularizer controlling the smoothness of the solution. Without regularization, the solution degenerates to $\operatorname{argmax}(z)$; all probability mass is assigned to the most probable item. For $\alpha = 1$ and Shannon entropy, the standard softmax transformation is recovered. Values $\alpha > 1$ produce sparse mappings in which part of the outputs become exactly zero. Increasing $\alpha$ gradually transitions the mapping toward more sparse solutions. For $\alpha=2$, the mapping reduces to sparsemax~\cite{martins2016softmax}.

Unlike softmax, sparse probability mappings cannot be directly substituted into the standard cross-entropy loss. Since entmax assigns zero probability to some items, the logarithm in the CE objective becomes undefined. Moreover, naive substitution does not preserve the desirable gradient structure. To address this issue, entmax is trained using the Fenchel-Young loss framework~\cite{blondel2019learning,blondel2020learning}. Given a prediction mapping $\mathbf{p}^*(\mathbf{z}) = \arg\max_{\mathbf{p} \in \Delta}
    \left\{
    \langle \mathbf{p}, \mathbf{z} \rangle -\Omega(\mathbf{p})
    \right\}$, the corresponding Fenchel-Young loss is defined as
\begin{equation}
\mathcal{L}_{\mathrm{FY}}(\mathbf{z},\mathbf{y}) = \Omega^*(\mathbf{z}) + \Omega(\mathbf{y}) - \langle \mathbf{z}, \mathbf{y} \rangle
\label{eq:fy_loss}
\end{equation}
where $\Omega$ is the regularization function and $\Omega^*$ is its convex conjugate $\Omega^*(\mathbf{z}) = \max_{\mathbf{p} \in \Delta}
    \left\{
    \langle \mathbf{p}, \mathbf{z} \rangle - \Omega(\mathbf{p})
    \right\}$. For entmax, $\Omega(\mathbf{p}) = -\mathcal{H}_\alpha(\mathbf{p})$. A key property of the Fenchel-Young framework is that the gradient of the loss preserves the familiar probabilistic form
\begin{equation}
\label{eq:gradient}
    \nabla_\mathbf{z} \mathcal{L}_{\mathrm{FY}}(\mathbf{z},\mathbf{y}) = \mathbf{p}^*(\mathbf{z}) - \mathbf{y},
\end{equation}
mirroring the gradient of softmax cross-entropy.

The main difference lies in sparsity. Only negative items with non-zero probability receive non-zero gradients. Thus, entmax restricts competition to a context-dependent subset of items. Items with sufficiently high logits remain inside the active support and compete with the target item, while low-scoring irrelevant items are excluded from optimization. In this sense, entmax can be interpreted as a differentiable adaptive hard negative selection mechanism: it focuses learning on difficult negatives determined by the current context rather than uniformly competing against the entire catalog.

The sparsity property of entmax improves training efficiency and ranking quality by restricting competition to contextually relevant alternatives, but does not directly address popularity bias. The active support is determined entirely by current logits, with no mechanism to account for global item popularity. Consequently, entmax can be expected to improve ranking accuracy on its own, while its effect on popularity-related metrics remains limited and dataset-dependent. We confirm this empirically in Section~\ref{sec:ablation}. Addressing popularity bias requires augmenting sparse mapping with explicit prior-awareness, which we develop next.

\section{Sparse Prior-Regularized Objective}

Our goal is to make sparse competition explicitly popularity-aware: popular items should be more likely to compete with the target when they are contextually plausible, while irrelevant items should still be filtered out by the sparse mapping. To achieve this, we incorporate prior information directly into the variational principle that defines the probability mapping. Figure~\ref{fig:illustration} shows how our approach affects model training.

\subsection{Prior-Regularized Variational Formulation}
\label{sec:pr_entmax}

We propose the following generalized variational problem to obtain a mapping from logits $\mathbf{z}$ to probabilities $\mathbf{p}$:
\begin{equation}
    \mathbf{p}^*(\mathbf{z}) =
    \arg\max_{\mathbf{p} \in \Delta}
    \left\{
    \langle \mathbf{p}, \mathbf{z} \rangle
    +
    \theta \, \mathcal{H}_\alpha(\mathbf{p})
    -
    \gamma \mathcal{H}(\mathbf{p},\mathbf{q})
    \right\}.
\label{eq:pr_entmax_variational}
\end{equation}
where, as in (4), $\Delta$ is the probability simplex, $\langle \mathbf{p}, \mathbf{z} \rangle$ denotes the dot product, and $\mathcal{H}_\alpha(\mathbf{p})$ is the Tsallis entropy (5).

\begin{equation}
\mathcal{H}(\mathbf{p},\mathbf{q}) = 
- \sum_{i \in \mathcal{I}} p_i \log q_i
\end{equation}
is the Shannon cross-entropy with respect to the prior item probabilities $q_i$. The prior term biases the learned distribution toward items with larger prior probabilities, while preserving the sparse support induced by Tsallis regularization. Hyperparameters $\theta$ and $\gamma$ control the strength of entropy regularization and prior regularization, respectively.

Applying the Karush--Kuhn--Tucker (KKT) conditions to the variational problem in (8) gives the following closed-form probability mapping (see Appendix~\ref{sec:appendix} for the full derivation):
\begin{equation}
    p_i^*(\mathbf{z})
    =
    \left[
    (\alpha-1)\left(\frac{z_i + \gamma \log q_i} {\theta}- \tau \right)
    \right]_+^{\frac{1}{\alpha - 1}},
\label{eq:pr_entmax_solution}
\end{equation}
where $[x]_+ = \max(0,x)$, and the threshold  $\tau$ is chosen such that probabilities sum to one:
\begin{equation}
\sum_{i \in \mathcal{I}} p_i^*(\mathbf{z})=1.
\end{equation}
As a result, prior information enters the sparse mapping additively in logit space through the correction term $\gamma \log q_i$. The threshold $\tau$ determines the active support of the distribution: items whose prior-adjusted logits fall below the threshold receive exactly zero probability. Consequently, the support of the resulting mapping is given by
\begin{equation}
\operatorname{supp}(\mathbf{p}^*) = 
\left\{
i: z_i + \gamma \log q_i >  \theta\tau
\right\}.
\label{eq:pr_support}
\end{equation}
This is a central property of the proposed formulation. The popularity prior affects not only the magnitude of the probabilities, but also which items belong to the active support and therefore participate in optimization. Only items with sufficiently large prior-adjusted logits receive non-zero probability and contribute to gradient updates.

Using the general Fenchel-Young construction from (\ref{eq:fy_loss}), the corresponding training objective can be written as
\begin{equation}
    \mathcal{L}_{_{\mathrm{PR-entmax}}}(\mathbf{z},\mathbf{y})
    = \langle \mathbf{p}^* - \mathbf{y}, \mathbf{z} \rangle +\theta \mathcal{H}_\alpha(\mathbf{p}^*) - \gamma \mathcal{H}(\mathbf{p}^*,\mathbf{q}),
\label{eq:pr_loss}
\end{equation}
where we omit the additive term $-\gamma \log q_y$, which is independent of the model parameters and does not affect the optimization.

Importantly, the Fenchel-Young framework preserves the standard probabilistic gradient structure from (\ref{eq:gradient}), and the proposed objective can be used as a drop-in replacement for softmax cross-entropy without modifying the standard training pipeline. The key difference is that the probability mapping is now both sparse and prior-aware.

The resulting training procedure can be interpreted as follows (see Figure~\ref{fig:illustration}). First, logits are adjusted using the additive prior correction $\gamma \log q_i$. The adjusted logits are then scaled by the parameter $\theta$, which acts as a temperature parameter in softmax-based models~\cite{agarwala2010temperature}. Finally, the sparse transformation in (\ref{eq:pr_entmax_solution}) is applied, and optimization is performed using the corresponding Fenchel-Young loss (\ref{eq:pr_loss}).

The final objective combines sparse context-dependent competition with explicit popularity-aware regularization. We therefore refer to the proposed approach as \textit{PR-entmax} (Prior-Regularized entmax).

\subsection{Score Decomposition and Inference}
\label{sec:score_decomposition}

The solution in (\ref{eq:pr_entmax_solution}) reveals that during training, the proposed objective operates on an adjusted score of the form
\begin{equation}
s_i^{\mathrm{train}}(x) = 
z_i(x)
+
\gamma \log q_i,
\label{eq:train_score}
\end{equation}
where $z_i(x)$ is the context-dependent model logit for item $i$ given the input sequence $x$, and $\gamma \log q_i$ is a global popularity-dependent correction term derived from the item prior. The training objective, therefore disentangles two different sources of preference signal: contextual item relevance and global item popularity. Since popularity is explicitly incorporated during training, the model learns logits that represent a contextual residual score beyond prior popularity.

At training time, these prior-adjusted scores determine which items enter the sparse support and compete with the target item. At inference time, however, we want to get a debiased contextual signal. We therefore remove the prior term and rank items using the raw model logits:
\begin{equation}
s_i^{\mathrm{inference}}(x) = z_i(x).
\label{eq:inference_score}
\end{equation}
In this way, popularity information is used during training to shape sparse competition, while the final ranking is determined solely by the learned contextual component.

\subsection{Connection to Existing Objectives}
\label{sec:connection}

The proposed formulation generalizes several existing output objectives as special cases. When $\gamma = 0$, the prior regularization term vanishes and (\ref{eq:pr_entmax_variational}) reduces to the entmax variational objective from (\ref{eq:entmax_mapping}). In the limit $\alpha \to 1$, Tsallis entropy converges to Shannon entropy, and the sparse mapping becomes dense softmax. When additionally $\gamma = 0$, the formulation reduces to the standard softmax cross-entropy objective. Finally, taking the dense limit $\alpha \to 1$ while retaining prior regularization ($\gamma > 0$) and setting $\theta=1$ yields
\begin{equation}
p_i(\mathbf{z})
\sim
\exp(z_i + \gamma \log q_i) =
q_i^\gamma \exp(z_i),
\end{equation}
which corresponds to softmax with additive log-prior correction, commonly known as logit adjustment~\cite{menon2020long}. Under our formulation, logit adjustment emerges naturally as the dense limit of a sparse prior-regularized objective, providing a variational interpretation and additional justification for additive popularity correction in logit space.

\begin{table}[h]
\centering
\caption{Comparison of dense and sparse output objectives.}
\label{tab:objective_comparison}
\small
\resizebox{\columnwidth}{!}{
\begin{tabular}{lccc}

\hline
\textbf{Objective} & \textbf{Sparse support} & \textbf{Prior correction} & \textbf{Prior affects support} \\
\hline
Cross-Entropy & No & No & No \\
Entmax & Yes & No & No \\
Logit Adjustment & No & Yes & No \\
PR-entmax & Yes & Yes & Yes \\
\hline

\end{tabular}
}
\end{table}

The key distinction between existing objectives and PR-entmax is summarized in Table~\ref{tab:objective_comparison}. Standard softmax cross-entropy induces dense competition without prior information. Entmax introduces sparse support, but the support remains prior-agnostic. Logit adjustment incorporates popularity priors, but the support remains dense. In contrast, the proposed objective provides sparse competition with prior-aware support selection.

\subsection{Mechanism: Context-Dependent Debiasing}
\label{sec:mechanism}

The proposed objective combines sparse competition and prior correction in a synergistic way. Unlike standard logit adjustment, the popularity prior affects not only the values of the probabilities, but also which items participate in the gradient updates through the sparse support condition (\ref{eq:pr_support}). Sparsity acts as a mechanism that controls the interaction between contextual relevance and prior correction. Popular but irrelevant items receive zero probability. Popular and relevant items remain in the competition with adjustment. Unpopular but highly relevant items can also enter the support through sufficiently strong contextual scores.

As a result, the model is trained to discriminate the target item from a dynamically selected set of hard negatives that are both contextually relevant and popularity-aware. This mechanism can be interpreted as a differentiable, context-dependent form of popularity-based negative sampling: the prior increases the chance that popular items enter the competition, but only when they are plausible alternatives.

To summarize, PR-entmax provides context-dependent debiasing through prior-aware support selection: popularity shapes the competitive structure of training, while final ranking is performed using the learned contextual scores.

\section{Experiments}

This section evaluates PR-entmax across multiple datasets and sequential recommendation architectures. We first describe the experimental setup and overall performance, and then analyze the contribution of sparsity and prior correction, the relevance–debiasing trade-off, exposure patterns, and sensitivity to the sparsity parameter.

\subsection{Experimental Settings}

\subsubsection{Datasets}
\begin{table}[!htbp]
\caption{\textbf{Statistics of the datasets after preprocessing.}}
\label{tab:datasetStats}
\centering
\resizebox{\columnwidth}{!}{%
    \begin{tabular}{lrrrr}
    \hline 
        \multirow{2}{*}{\textbf{Dataset}} & \multirow{2}{*}{\textbf{\# Users}} & \multirow{2}{*}{\textbf{\# Items}} & \multirow{2}{*}{\textbf{\# Interact.}} & \multicolumn{1}{c}{\textbf{Avg.}} \\ 
        & & & & \multicolumn{1}{c}{\textbf{length}}\\ \hline     
        \textbf{ML-1M \cite{harper2015movielens}}  & 6,040 & 3,373 & 999,295 & 165 \\

        \textbf{Tmall} \footnotemark[1] & 44,759 & 173,427 & 3,721,625 & 83 \\
        \textbf{Yambda-50M \cite{ploshkin2025yambda}} & 9,006 & 38,399 & 888,332 & 99 \\

        \textbf{Yoochoose \cite{ben2015recsys}} & 142,954 & 16,068 & 1,212,835 & 9  \\

        \textbf{Zvuk \cite{shevchenko2024variability}} & 9,272 & 109,346 & 4,695,928 & 506\\

    \hline 
    \end{tabular}
}
\end{table}

\footnotetext[1]{\url{https://tianchi.aliyun.com/dataset/42}}

We conduct experiments on five datasets. \break MovieLens-1M (ML-1M) is a widely used benchmark for recommendation systems in the movie domain. Tmall and Yoochoose are e-commerce datasets commonly used in sequential recommendation research ~\cite{klenitskiy2024does}. Zvuk and Yambda-50M are music streaming datasets with strong sequential patterns~\cite{klenitskiy2026analysis}.
For all datasets, we apply identical $n$-core filtering with $n=5$  to the training set. Additionally, for Zvuk and Tmall, we randomly sample 10,000 and 50,000 users, respectively, to obtain computationally efficient yet representative subsets. For Yambda-50M, we truncate user interaction sequences to a maximum length of 128 prior to preprocessing, as the original dataset contains very long user histories. Table~\ref{tab:datasetStats} contains the main statistics of the datasets after preprocessing.

\subsubsection{Evaluation}
We use a global temporal split~\cite{ji2023critical,sun2023take,gusak2025time} where the cutoff is placed at 90\% of the interactions. The validation set is constructed by randomly sampling 10\% of users from the training data. For both validation and test sets, we treat the last user interaction as the prediction target, while all preceding interactions form the input sequence.

Recommendation quality is evaluated using NDCG@k with $k \in \{10,100\}$. To quantify popularity bias, we report two complementary metrics. Coverage@k measures overall item exposure, while Novelty@k (sometimes also referred to as Surprisal@k) measures the average normalized self-information of recommended items, assigning higher scores to less popular items.
We use its implementation from the Replay library~\cite{vasilev2024replay}. For each item $j$, novelty is defined as:
\begin{equation}
    \operatorname{Novelty}(j)
    = \frac{- \log_2(u_j/|\mathcal{U}|)}{\log_2{|\mathcal{U}|}}
    = 1 - \frac{\log_2{ u_j}}{\log_2{|\mathcal{U}|}},
\end{equation}
where $u_j$ is the number of users who interacted with item $j$, and $|\mathcal{U}|$ is the total number of users in train set. For cold items, we set $u_j = 1$. The final metric is averaged over the top-$k$ recommendation lists and then over users in test set:
\begin{equation}
    \operatorname{Novelty}@k
    =
    \frac{1}{|\mathcal{U}_{test}|}
    \sum_{i \in \mathcal{U}_{test}}
    \frac{1}{k}
    \sum_{j \in R_i^k}
    \operatorname{Novelty}(j),
\end{equation}
where $R_i^k$ denotes the top-$k$ recommendation list for user $i$.

\subsubsection{Baselines.}

The natural baselines for comparison with PR-entmax are approaches that also modify the training objective to mitigate popularity bias. Accordingly, we include several methods that alter the loss formulation.

\begin{itemize}
\item \textit{Cross-Entropy} \textit{(CE)}  loss with softmax is the primary baseline for comparison for PR-enmtax and all other approaches, since it is widely adopted in sequential recommendation models, and we build other baselines on top of it.
\item \textit{IPW}: \textit{Inverse Propensity Weighting}~\cite{schnabel2016recommendations} re-weights the standard cross-entropy loss by down-weighting popular items during training. The objective is defined as:
    \begin{equation}
        \mathcal{L}_{\text{\textit{IPW}}}(x, y) = - \frac{1}{n_{y}^{\gamma}} \log \frac{\exp(z_y(x))}{\sum_{j \in I} \exp(z_j(x))}
    \end{equation}
where $n_{y}$ denotes the frequency of the target item $y$, smoothed with parameter $\gamma$.
\item \textit{PD}: \textit{Popularity Deconfounding} \cite{zhang2021causal} modifies item logits during training by explicitly incorporating prior item popularity probabilities. Following the original implementation, we first transform logits with a positive ELU-based mapping and then scale them by the item prior:
    \begin{equation}
        \widetilde{z}_j(x) =
        \left(\operatorname{ELU}(z_j(x)) + 1\right)
        \cdot q_j^{\gamma}.
    \end{equation}
The final objective is the standard cross-entropy loss computed over the adjusted logits:
    \begin{equation}
        \mathcal{L}_{\text{\textit{PD}}}(x, y)
        =
        - \log
        \frac{\exp(\widetilde{z}_y(x))}
        {\sum_{j \in I} \exp(\widetilde{z}_j(x))}.
    \end{equation}
Here $q_j$ denotes the prior probability of item $j$ estimated from item popularity, and $\gamma$ controls the strength of the popularity adjustment.

\item \textit{Pop Sampling}:  \textit{Popularity-based Negative sampling} \cite{prakash2024evaluating} \break
changes the sampling distribution of negative items according to prior item probabilities while keeping the standard cross-entropy objective. 
    \begin{equation}
        p_{\text{neg}}(j)
        =
        \frac{q_j^{\gamma}}
        {\sum_{k \in I 
        } q_k^{\gamma}},
        \qquad j \neq y.
    \end{equation}
Let $N_y$ be a set of negative items of fixed size $N$ sampled according to $p_{\text{neg}}$. The loss is defined as:
    \begin{equation}
        \mathcal{L}_{\text{\textit{Pop Sampling}}}(x, y)
        =
        - \log
        \frac{\exp(z_y(x))}
        {\sum_{j \in N_y} \exp(z_j(x))}.
    \end{equation}
\item \textit{Logit Adjustment}\cite{menon2020long}: modifies the logits before applying the cross-entropy loss by adding a popularity-dependent bias term based on prior item probabilities. The adjusted logits are defined as:
    \begin{equation}
        \widetilde{z}_j(x)
        =
        z_j(x) + \gamma \log q_j .
    \end{equation}
The objective is then:
    \begin{equation}
        \mathcal{L}_{\text{\textit{LA}}}(x, y)
        =
        - \log
        \frac{\exp(\widetilde{z}_y(x))}
        {\sum_{j \in I} \exp(\widetilde{z}_j(x))}.
    \end{equation}
Here $q_j$ denotes the prior probability of item $j$ estimated from the training data, and $\gamma$ controls the strength of the logit adjustment.
\end{itemize}

\subsubsection{Implementation Details}

We conduct experiments using two widely used sequential recommendation models: SASRec~\cite{kang2018self} and GRU4Rec~\cite{hidasi2015session}. 
Hyperparameters are tuned separately for each method and backbone. 
For SASRec, we tune the hidden size, the number of self-attention blocks, the dropout rate, and the number of attention heads. 
For GRU4Rec, we tune the hidden dimensionality, the number of recurrent layers, and the dropout rate. 
For the Entmax loss, we additionally tune the temperature parameter $\theta$, which controls sparsity and is critical for strong performance. 
In contrast, tuning the temperature for softmax does not yield significant improvements.

Since the remaining methods are based on either Cross-Entropy or Entmax, we first identify the best hyperparameters for these base losses and then tune the method-specific parameter $\gamma$ for  PR-entmax, Logit Adjustment, Pop Sampling, Popularity Deconfounding, and Inverse Propensity Weighting. For Pop Sampling, we additionally tune the number of sampled negative items $N_{y}$. We choose $N_{y}$ large enough to ensure that the quality degradation is not caused merely by using too few negative samples.

All models are trained with a maximum input sequence length of 128 and a batch size of 64, using the Adam optimizer with a learning rate of 1e-3. For Tmall, due to the large number of items, we apply random negative sampling in the loss function with 50,000 negative examples. To improve statistical significance, we run each experiment three times with different random seeds and report averaged metrics.

In our main experiments, we use Entmax and PR-entmax with $\alpha=1.5$. We analyze the effect of varying $\alpha$ in Section~\ref{sec:sparsity} and find that this is a good choice for all datasets. Additionally, for this case, there exists an effective algorithm for probability mapping~\cite{peters2019sparse}.

The source code and detailed hyperparameter settings are provided in the GitHub repository\footnotemark[2].

\footnotetext[2]{\url{https://github.com/Antondfger/Sparse-Prior-Regularized/tree/cikm_2026}}

\subsection{Overall Performance}
\label{sec:overall}

\begin{table}[t]

\caption{Experimental results on SASRec for all considered approaches. Bold numbers mark the best model. The second best is underlined. Asterisk ($^{*}$) denotes the statistically significant improvement over \textit{Cross-Entropy} for the best model.}
\label{tab:results_SASRec}
\centering
\resizebox{\linewidth}{!}{%
\setlength{\tabcolsep}{4pt} %
\small
\begin{tabular}{@{}llcccccc@{}}
\toprule
\textbf{Dataset} & \textbf{Method} & \multicolumn{2}{c}{\textbf{NDCG}} & \multicolumn{2}{c}{\textbf{Novelty}} & \multicolumn{2}{c}{\textbf{Coverage}} \\
\cmidrule(lr){3-4} \cmidrule(lr){5-6} \cmidrule(lr){7-8}
 & & @10 & @100 & @10 & @100 & @10 & @100 \\
\midrule
\multirow{4}{*}{ML1M} 
& \textit{Cross-Entropy} & 0.047 & \underline{0.102} & 0.274 & 0.308 & 0.463 & 0.835 \\

& \textit{IPW} & 0.047 & 0.100 & 0.290 & 0.321 & 0.496 & 0.871  \\

& \textit{PD} & 0.040 & 0.095 & \underline{0.306} & \underline{0.327} & \underline{0.514} & 0.851  \\

& \textit{Pop Sampling} & \underline{0.049} & \underline{0.102} & 0.286 & 0.319 & 0.489 & \underline{0.872}  \\

& \textit{Logit Adjustment} & 0.047 & 0.100 & 0.291 & 0.321 & 0.496 & 0.869 \\

\arrayrulecolor{gray!80}
\cline{2-8}
\arrayrulecolor{black}

& \textit{PR-entmax} & \textbf{0.058}$^{*}$  & \textbf{0.103} & \textbf{0.331}$^{*}$  & \textbf{0.411}$^{*}$  & \textbf{0.600}$^{*}$  & \textbf{0.993}$^{*}$  \\

\midrule
\multirow{4}{*}{Tmall} 
& \textit{Cross-Entropy} & \underline{0.274} & \underline{0.305} & 0.657 & 0.664 & 0.351 & 0.649 \\
& \textit{IPW} & 0.265	& 0.296 & 0.704 &	0.704 & 	0.431 &	0.722 \\

& \textit{PD} & 0.249 & 0.279 & 0.705 & 0.713 & 0.332 & 0.566 \\

& \textit{Pop Sampling} & 0.272 & 0.301 & \textbf{0.729}$^{*}$ & \textbf{0.755}$^{*}$ & \underline{0.432} & \underline{0.758} \\

& \textit{Logit Adjustment} & 0.271	& 0.301	& \underline{0.711}	& \underline{0.733}	& 0.416 & 0.741\\

\arrayrulecolor{gray!80}
\cline{2-8}
\arrayrulecolor{black}

& \textit{PR-entmax} & \textbf{0.285}$^{*}$ & \textbf{0.312}$^{*}$ & 0.703 & 0.730 & \textbf{0.499}$^{*}$ & \textbf{0.878}$^{*}$ \\
\midrule

\multirow{4}{*}{Yambda-50M}
& \textit{Cross-Entropy} & 0.089 & \underline{0.118} & 0.553 & 0.562 & 0.440 & 0.781 \\
& \textit{IPW} & 0.085 & 0.106 & \textbf{0.687}$^{*}$ & \underline{0.682} &  \underline{0.696} & 0.959 \\

& \textit{PD} & 0.084 & 0.110 & 0.617 & 0.620 & 0.472 & 0.731\\

& \textit{Pop Sampling} & \underline{0.092} & 0.116 & 0.647 & 0.658 & 0.668 & \underline{0.960}\\

& \textit{Logit Adjustment} & 0.089 & 0.113 & 0.626 & 0.637 & 0.550 & 0.878 \\

\arrayrulecolor{gray!80}
\cline{2-8}
\arrayrulecolor{black}

& \textit{PR-entmax} & \textbf{0.098}$^{*}$ & \textbf{0.120}$^{*}$ & \underline{0.671} & \textbf{0.688}$^{*}$ & \textbf{0.785}$^{*}$ & \textbf{0.997}$^{*}$ \\
\midrule
\multirow{4}{*}{Yoochoose}
& \textit{Cross-Entropy} & \underline{0.203} & \textbf{0.230} & 0.606 & 0.609 & 0.753 & 0.987 \\
& \textit{IPW} & 0.191 & 0.219 & \underline{0.655} & 0.680 & 0.832 & \textbf{0.999} \\

& \textit{PD} & 0.188 & 0.216 & 0.645 & 0.655 & 0.768 & 0.977 \\

& \textit{Pop Sampling} & 0.200 & 0.226 & \textbf{0.663}$^{*}$ & \underline{0.695} & \underline{0.866} & \underline{0.998} \\

& \textit{Logit Adjustment} & 0.194 & 0.221 & 0.653 & 0.683 & 0.839 & \underline{0.998} \\

\arrayrulecolor{gray!80}
\cline{2-8}
\arrayrulecolor{black}

& \textit{PR-entmax} & \textbf{0.209}$^{*}$ & \underline{0.228} & 0.646 & \textbf{0.755}$^{*}$ & \textbf{0.876}$^{*}$ & \textbf{0.999}$^{*}$ \\
\midrule

\multirow{4}{*}{Zvuk}
& \textit{Cross-Entropy} & \underline{0.218} & \textbf{0.256} & 0.550 & 0.539 & 0.163 & 0.462 \\
& \textit{IPW} & 0.199 & 0.228 & \underline{0.655} & 0.661 & 0.182 & 0.436 \\

& \textit{PD} & 0.199 & 0.237 & 0.618 & 0.616 & 0.176 & 0.425\\

& \textit{Pop Sampling} & 0.212 & 0.242 & 0.610 & \underline{0.668} & 0.150 & 0.397\\

& \textit{Logit Adjustment} & 0.214 & \underline{0.250} & 0.637 & 0.657 & \underline{0.205} & \underline{0.559} \\

\arrayrulecolor{gray!80}
\cline{2-8}
\arrayrulecolor{black}

& \textit{PR-entmax} & \textbf{0.222}$^{*}$ & 0.246 & \textbf{0.686}$^{*}$ & \textbf{0.806}$^{*}$ & \textbf{0.216}$^{*}$ & \textbf{0.579}$^{*}$ \\
\bottomrule
\end{tabular}
}
\end{table}

\begin{table}[t]

\caption{Experimental results on GRU4Rec for all considered approaches. Bold numbers mark the best model. The second best is underlined.  Asterisk ($^{*}$) denotes the statistically significant improvement over \textit{Cross-Entropy} for the best model.}
\label{tab:results_GRU4Rec}
\centering
\resizebox{\linewidth}{!}{%
\setlength{\tabcolsep}{4pt} %
\small
\begin{tabular}{@{}llcccccc@{}}
\toprule
\textbf{Dataset} & \textbf{Method} & \multicolumn{2}{c}{\textbf{NDCG}} & \multicolumn{2}{c}{\textbf{Novelty}} & \multicolumn{2}{c}{\textbf{Coverage}} \\
\cmidrule(lr){3-4} \cmidrule(lr){5-6} \cmidrule(lr){7-8}
 & & @10 & @100 & @10 & @100 & @10 & @100 \\
\midrule
\multirow{4}{*}{ML1M} 
& \textit{Cross-Entropy} & \underline{0.054} & \textbf{0.105} & 0.258 & 0.287 & 0.463 & 0.858 \\
& \textit{IPW} & 0.048 & \underline{0.100} & \textbf{0.308}$^{*}$ & \underline{0.327} & \underline{0.599} & \underline{0.952}  \\

& \textit{PD} & 0.045 & 0.095 & \underline{0.303} & 0.318 & 0.546 & 0.913\\

& \textit{Pop Sampling} & 0.050 & 0.099 & 0.265 & 0.290 & 0.487 & 0.872 \\

& \textit{Logit Adjustment} & \underline{0.054} & \textbf{0.105} & 0.290 & 0.313 & 0.555 & 0.928 \\

\arrayrulecolor{gray!80}
\cline{2-8}
\arrayrulecolor{black}

& \textit{PR-entmax} & \textbf{0.062}$^{*}$ & \underline{0.100} & 0.295 & \textbf{0.376}$^{*}$ & \textbf{0.616}$^{*}$ & \textbf{0.997}$^{*}$ \\
\midrule

\multirow{4}{*}{Tmall} 
& \textit{Cross-Entropy} & 0.216 & 0.241 & 0.654 & 0.672 & 0.377 & 0.727 \\
& \textit{IPW} &0.190 & 0.210 & 0.713 & 0.713 & 0.424 & 0.733 \\

& \textit{PD} & 0.205 & 0.229 & 0.710 & 0.721 & 0.445 & 0.762  \\

& \textit{Pop Sampling} & \underline{0.223} & \underline{0.245} & 0.749 & 0.769 & 0.480 & 0.805  \\

& \textit{Logit Adjustment} & 0.212 & 0.235 & \textbf{0.759}$^{*}$ & \underline{0.772} & \underline{0.492} & \underline{0.808}  \\

\arrayrulecolor{gray!80}
\cline{2-8}
\arrayrulecolor{black}

& \textit{PR-entmax} & \textbf{0.291}$^{*}$ & \textbf{0.313}$^{*}$ & \underline{0.751} & \textbf{0.795}$^{*}$ & \textbf{0.526}$^{*}$ & \textbf{0.897}$^{*}$ \\
\midrule

\multirow{4}{*}{Yambda-50M}
& \textit{Cross-Entropy} & 0.056 & \underline{0.085} & 0.578 & 0.589 & 0.145 & 0.605 \\
& \textit{IPW} & 0.054 & 0.082 & 0.642 & 0.641 & 0.666 & 0.985 \\

& \textit{PD} &  0.054 & 0.081 & \underline{0.651} & \underline{0.645} & \underline{0.668} & 0.980 \\

& \textit{Pop Sampling} & \underline{0.057} & \underline{0.085} & 0.633 & 0.643 & 0.608 & 0.970 \\

& \textit{Logit Adjustment} & 0.055 & \underline{0.085} & 0.632 & 0.636 & 0.629 & \underline{0.986} \\

\arrayrulecolor{gray!80}
\cline{2-8}
\arrayrulecolor{black}

& \textit{PR-entmax} & \textbf{0.086}$^{*}$ & \textbf{0.103}$^{*}$ &\textbf{ 0.725}$^{*}$ & \textbf{0.754}$^{*}$ &\textbf{ 0.738}$^{*}$ & \textbf{0.996}$^{*}$\\
\midrule
\multirow{4}{*}{Yoochoose}
& \textit{Cross-Entropy} & \textbf{0.199} & \textbf{0.225} & 0.598 & 0.605 & 0.734 & 0.987 \\

& \textit{IPW} & 0.195 & 0.221 & 0.632 & 0.658 & 0.817 & \underline{0.998} \\

& \textit{PD} & 0.180 & 0.208 & 0.648 & 0.666 & 0.785 & 0.978\\

& \textit{Pop Sampling} & \underline{0.197} & \underline{0.222} & \underline{0.656} & \underline{0.701} & 0.822	& 0.997\\

& \textit{Logit Adjustment} & 0.192& 0.218 & 0.648 & 0.684 & \underline{0.831} & \textbf{0.999}$^{*}$ \\

\arrayrulecolor{gray!80}
\cline{2-8}
\arrayrulecolor{black}

& \textit{PR-entmax} & \underline{0.197} & 0.213 & \textbf{0.696}$^{*}$ & \textbf{0.791}$^{*}$ & \textbf{0.884}$^{*}$ & 0.997 \\
\midrule

\multirow{4}{*}{Zvuk}
& \textit{Cross-Entropy} & 0.163 & 0.196 & 0.568 & 0.584 & 0.054 & 0.272 \\
& \textit{IPW} & 0.143 & 0.177 & \underline{0.619} & 0.626 & 0.187 & 0.529  \\

& \textit{PD} & 0.144 & 0.175 & \textbf{0.631}$^{*}$ & 0.641 & 0.181 & 0.520\\

& \textit{Pop Sampling} & \underline{0.165} & \underline{0.200} & 0.618 & \underline{0.654} & 0.183 & 0.524 \\

& \textit{Logit Adjustment} & 0.161 & 0.195 & 0.616 & 0.632 & \underline{0.188} & \textbf{0.546}$^{*}$ \\

\arrayrulecolor{gray!80}
\cline{2-8}
\arrayrulecolor{black}

& \textit{PR-entmax} & \textbf{0.224}$^{*}$ & \textbf{0.255}$^{*}$ & 0.612 & \textbf{0.667}$^{*}$ & \textbf{0.189}$^{*}$ & \underline{0.533} \\
\bottomrule
\end{tabular}
}
\end{table}

Table~\ref{tab:results_SASRec} reports the overall performance of all methods across five datasets for the SASRec backbone, while Table~\ref{tab:results_GRU4Rec} presents the corresponding results for GRU4Rec. Overall, PR-entmax consistently improves over the standard cross- entropy objective in both recommendation relevance and popularity debiasing metrics and achieves the strongest overall relevance-debiasing trade-offs among the compared methods.

Other methods improve Novelty and Coverage at the cost of a moderate degradation in NDCG relative to the Cross-Entropy baseline in most cases. In contrast, PR-entmax preserves competitive ranking quality and, in many cases, further improves it. NDCG gains are particularly pronounced for GRU4Rec, where it substantially improves ranking quality on all datasets except for Yoochoose, indicating that the sparse objective is especially beneficial for this type of model. At the same time, the proposed method also demonstrates strong popularity debiasing performance.

PR-entmax obtains the best results in terms of Coverage and Novelty@100 across all datasets and both backbones. For Novelty@10, the results depend on the dataset and model architecture. In particular, the proposed method is outside the top-2 methods on Tmall and Yoochoose for SASRec, and on ML-1M and Zvuk for GRU4Rec. Nevertheless, even in these cases, it substantially outperforms the standard Cross-Entropy baseline. In terms of ranking accuracy, PR-entmax achieves the best NDCG@10 results on all datasets except Yoochoose. For NDCG@100, the method remains highly competitive, being slightly below the strongest baseline on Yoochoose and Zvuk for SASRec, and on ML-1M and Yoochoose for GRU4Rec.

Overall, the observed improvements are consistent across both backbone models, indicating that the effect of prior-aware sparse competition generalizes across different sequential recommendation architectures.

Beyond recommendation quality, we verify that PR‑entmax introduces no additional time or memory overhead compared to standard Entmax. The prior term only adds a constant shift to the logits, leaving the sparse support computation unchanged. On large datasets (Zvuk), entmax‑based training is about 20\% slower than softmax cross‑entropy due to iterative computation of threshold $\tau$ that determines the sparse support. On smaller datasets (ML‑1M, Yoochoose), all methods are comparable. Importantly, PR‑entmax matches Entmax's runtime exactly, inheriting its scalability without extra cost.

\subsection{Ablation Study}
\label{sec:ablation}

\begin{figure*}[ht]
    \centering
    \begin{subfigure}{0.49\linewidth}
    \includegraphics[width=\textwidth]{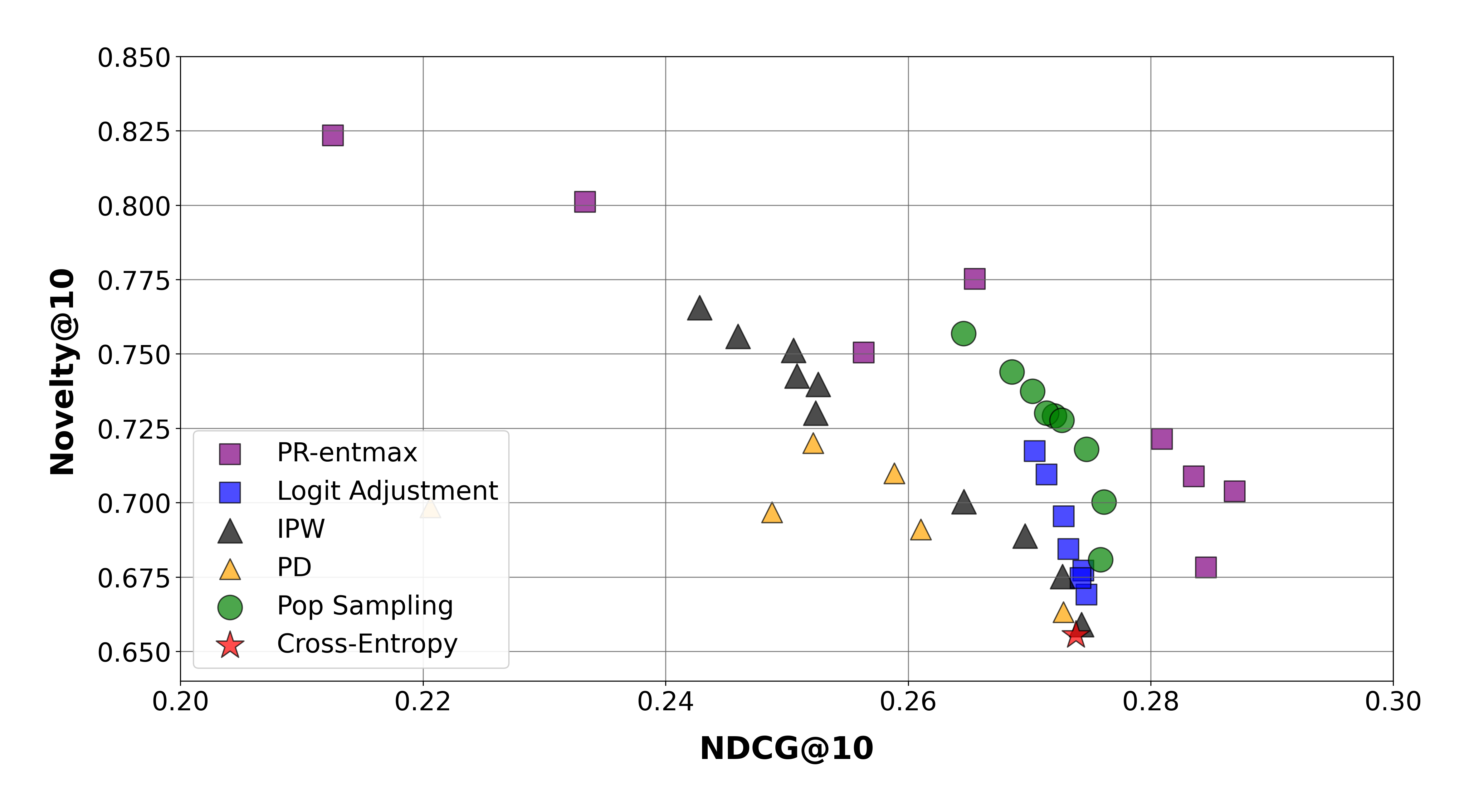}
    \caption{Tmall}
    \end{subfigure}
    \begin{subfigure}{0.49\linewidth}
    \includegraphics[width=\textwidth]{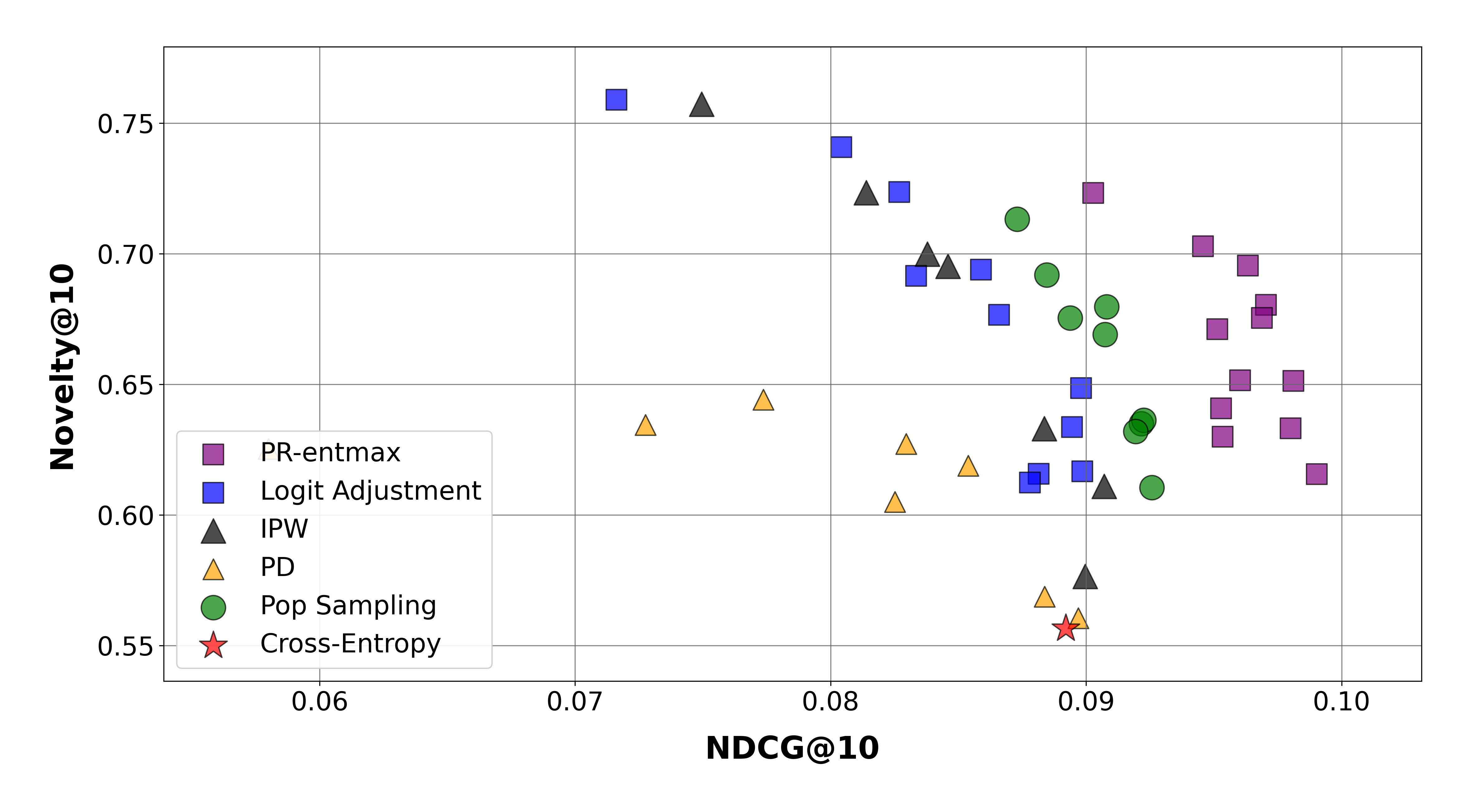}
    \caption{Yambda-50M}
    \end{subfigure}
    \begin{subfigure}{0.49\linewidth}
    \includegraphics[width=\textwidth]{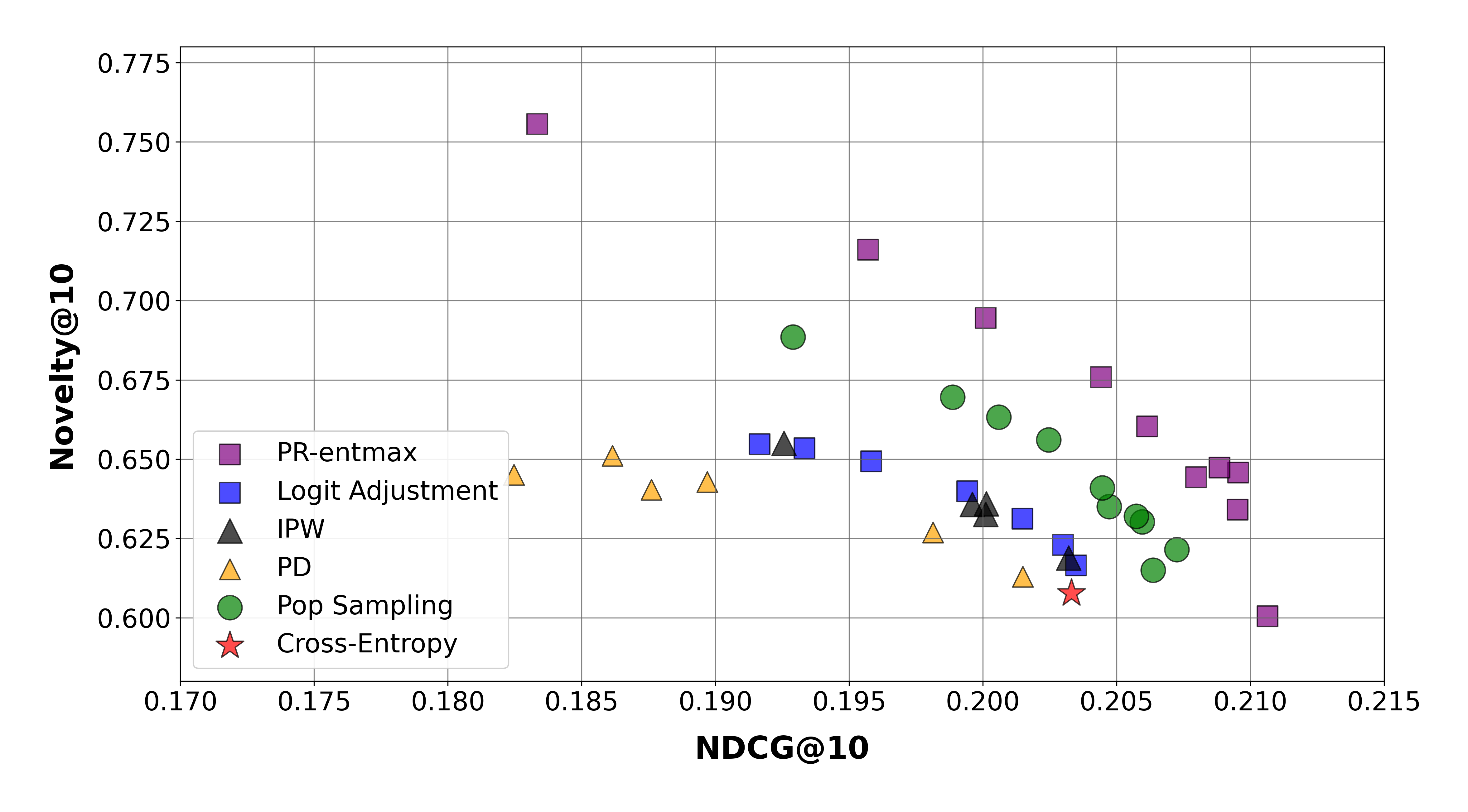}
    \caption{Yoochoose}
    \end{subfigure}
    \begin{subfigure}{0.49\linewidth}
    \includegraphics[width=\textwidth]{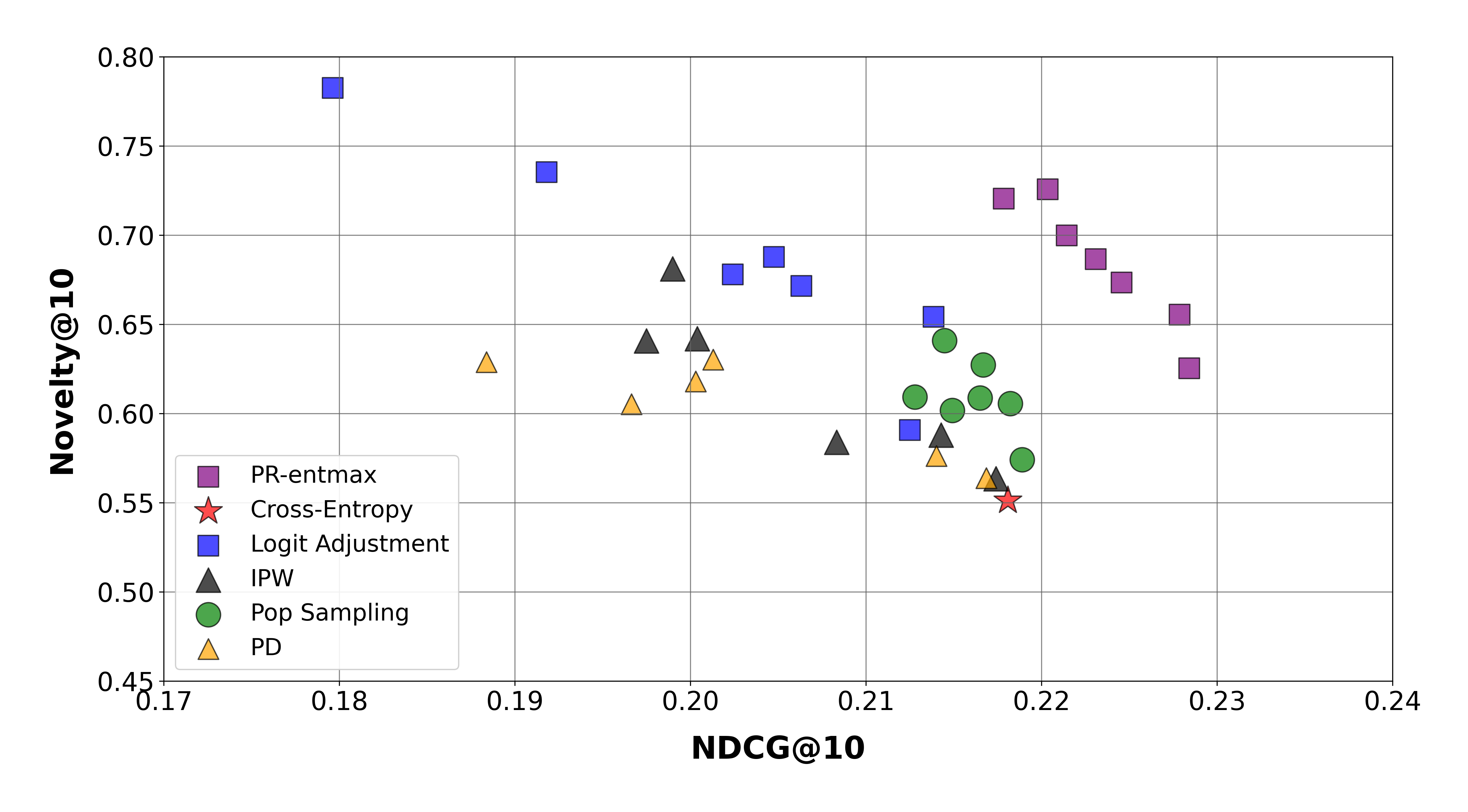}
    \caption{Zvuk}
    \end{subfigure}
    \caption{Trade-off between NDCG@10 and Novelty@10: Pareto-optimal fronts of losses across different datasets. The top-right direction indicates better performance in both metrics.}
    \label{fig:pareto}
\end{figure*}

\begin{table}[t]
\caption{Ablation study with relative improvements (\%) of different methods over Cross-Entropy baseline for SASRec. Bold numbers mark the best model.}
\label{tab:ablation}
\centering
\resizebox{\linewidth}{!}{%
\setlength{\tabcolsep}{4pt}
\small
\begin{tabular}{@{}llcccc@{}}
\toprule
\textbf{Dataset} & \textbf{Method} & \textbf{NDCG@10} & \textbf{Novelty@10} & \textbf{Coverage@10} \\
\midrule
\multirow{3}{*}{ML1M} 
& PR-entmax & \textbf{+18\%} & \textbf{+17\%} & \textbf{+23\%} \\
& Logit Adjustment & -1\% & +6\% & +6\% \\
& Entmax & \textbf{+18\%} & -7\% & -5\%\\
\midrule
\multirow{3}{*}{Tmall} 
& PR-entmax & \textbf{+4\%} & +7\% & \textbf{+30\%} \\
& Logit Adjustment & -5\% & \textbf{+8\%} & +18\% \\
& Entmax & \textbf{+4\%} & +4\% & +25\% \\
\midrule
\multirow{3}{*}{Yambda-50M}
& PR-entmax & \textbf{+9\%} & \textbf{+18\%} & \textbf{+44\%} \\
& Logit Adjustment & -1\% & +12\% & +20\% \\
& Entmax &\textbf{ +9\%} & +9\% & +34\% \\
\midrule
\multirow{3}{*}{Yoochoose}
& PR-entmax & +3\% & +6\% & \textbf{+14\%} \\
& Logit Adjustment & -5\% & \textbf{+7\%} & +10\% \\
& Entmax & \textbf{+4\%} & -1\% & +2\% \\
\midrule
\multirow{3}{*}{Zvuk}
& PR-entmax & +2\% &\textbf{ +20\%} & \textbf{+24\%} \\
& Logit Adjustment & -2\% & +14\% & +20\% \\
& Entmax &\textbf{ +6\%} & +9\% & +15\%\\
\bottomrule
\end{tabular}
}
\end{table}

To analyze the individual contributions of sparsity and prior correction, Table~\ref{tab:ablation} reports the relative improvements over the standard Cross-Entropy baseline for SASRec. We compare the proposed approach with Entmax, which introduces sparse competition without prior-awareness ($\gamma = 0$), and Logit Adjustment, which performs prior correction without sparsity (the dense limiting case corresponding to $\alpha \to 1$ and $\gamma > 0$).

Introducing sparsity through Entmax consistently improves ranking quality across all datasets, confirming the benefits of sparse probability mapping for recommendation relevance. However, its debiasing performance strongly depends on the dataset. In particular, Entmax does not improve debiasing metrics on Yoochoose and even reduces them on ML-1M. While it increases Novelty@10 and Coverage@10 on Tmall, Yambda-50M, and Zvuk, these gains remain substantially smaller than those achieved by prior-aware methods.

In contrast, Logit Adjustment consistently improves Novelty@10 and Coverage@10, indicating that explicit popularity-aware correction effectively shifts recommendation exposure toward less popular items. However, these improvements typically come at the cost of reduced ranking accuracy or only marginal relevance gains.

PR-entmax combines the strengths of both components. The proposed objective preserves the high-ranking performance associated with sparse mapping while simultaneously outperforming both Entmax and Logit Adjustment in terms of debiasing metrics.

\subsection{Relevance–Debiasing Trade-off}
\label{sec:tradeoff}

\begin{figure*}[!h]
    \centering
    \begin{subfigure}{0.24\linewidth}
    \includegraphics[width=\textwidth]{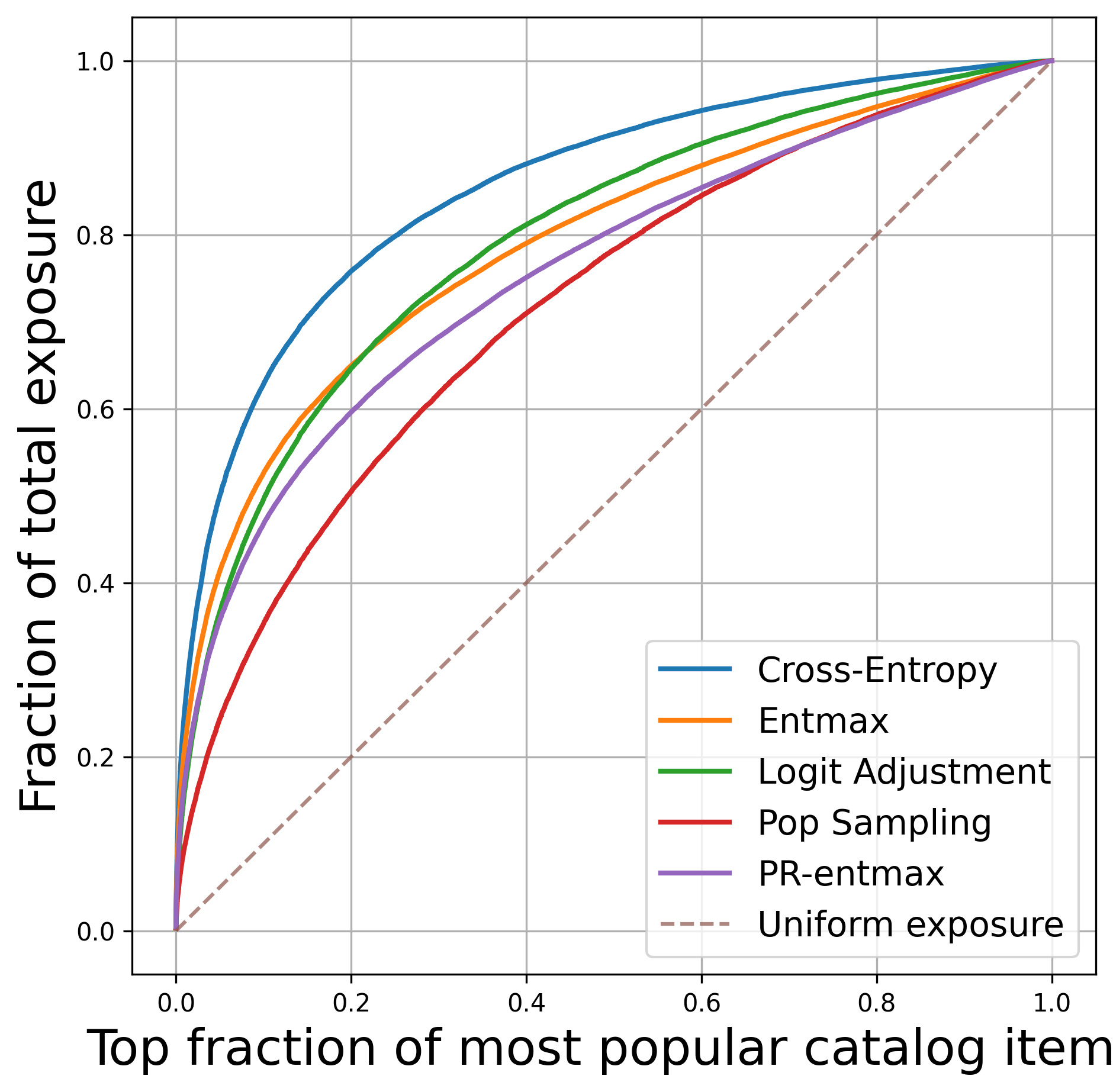}
    \caption{Tmall}
    \end{subfigure}
    \begin{subfigure}{0.24\linewidth}
    \includegraphics[width=\textwidth]{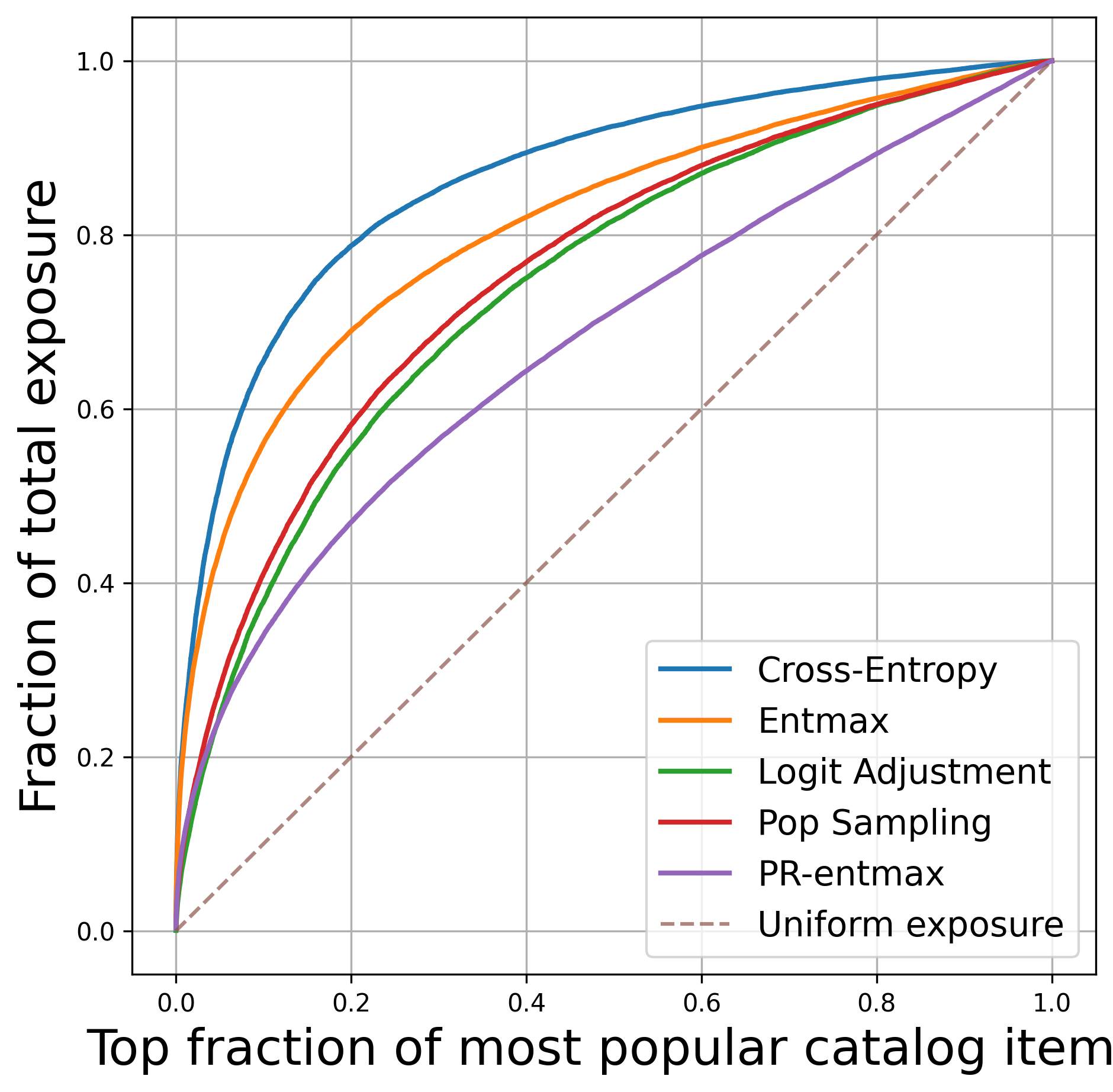}
    \caption{Yambda-50M}
    \end{subfigure}
    \begin{subfigure}{0.24\linewidth}
    \includegraphics[width=\textwidth]{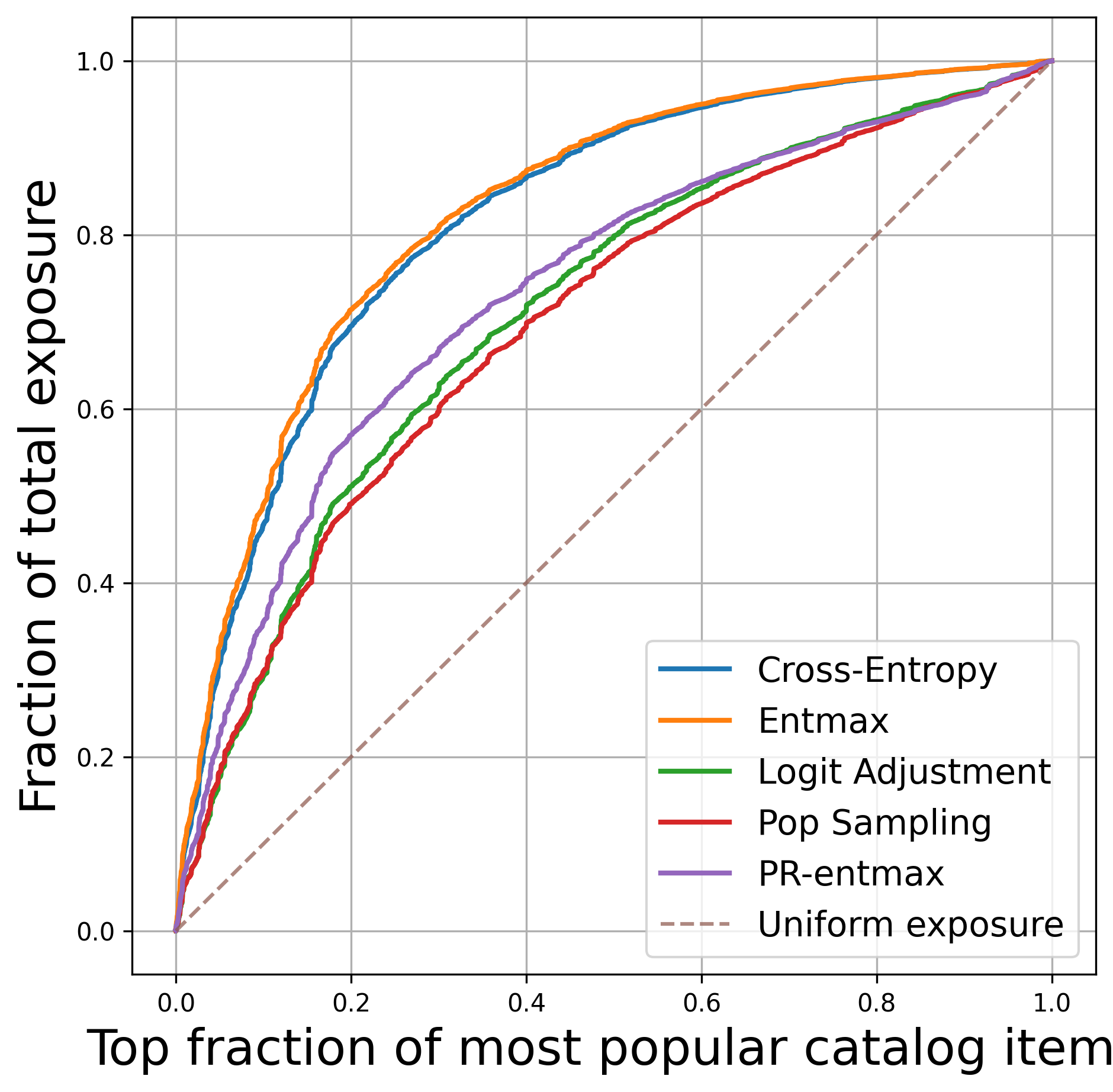}
    \caption{Yoochoose}
    \end{subfigure}
    \begin{subfigure}{0.24\linewidth}
    \includegraphics[width=\textwidth]{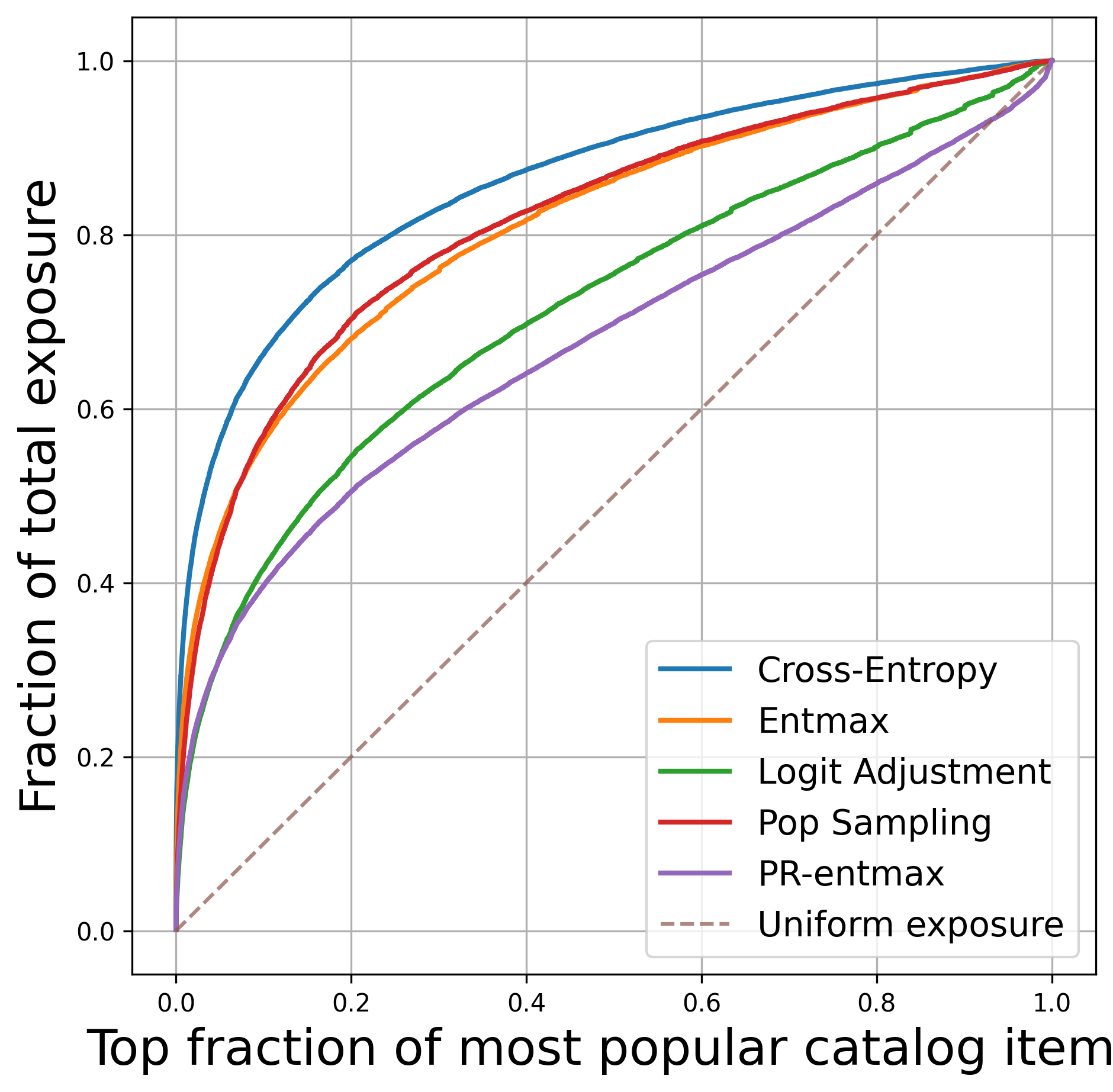}
    \caption{Zvuk}
    \end{subfigure}
    \caption{Exposure concentration of most popular items across different datasets. The x‑axis denotes the proportion of the most popular items, and the y‑axis shows the exposure assigned to these items. Lower curves indicate lower popularity bias, as they concentrate less exposure on popular items.}
    \label{fig:exposure}
\end{figure*}

\begin{figure*}[!h]
    \centering
    \begin{subfigure}{0.49\linewidth}
    \includegraphics[width=\textwidth]{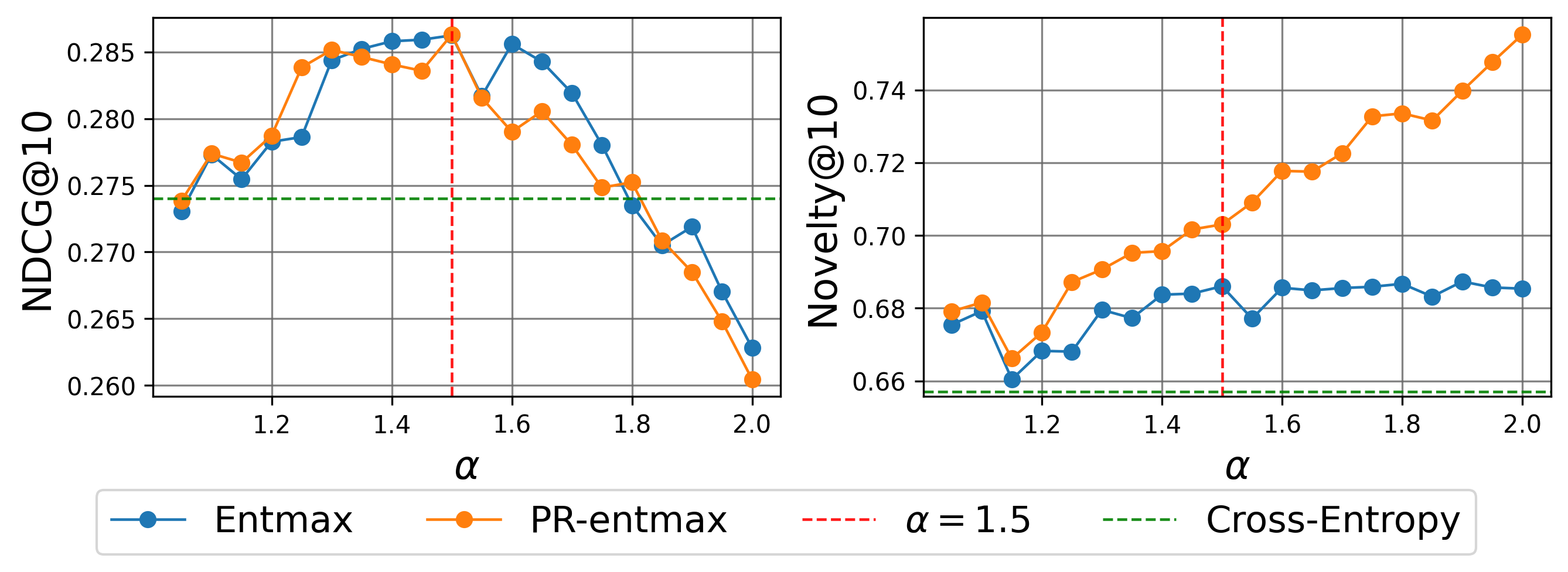}
    \caption{Tmall}
    \end{subfigure}
    \begin{subfigure}{0.49\linewidth}
    \includegraphics[width=\textwidth]{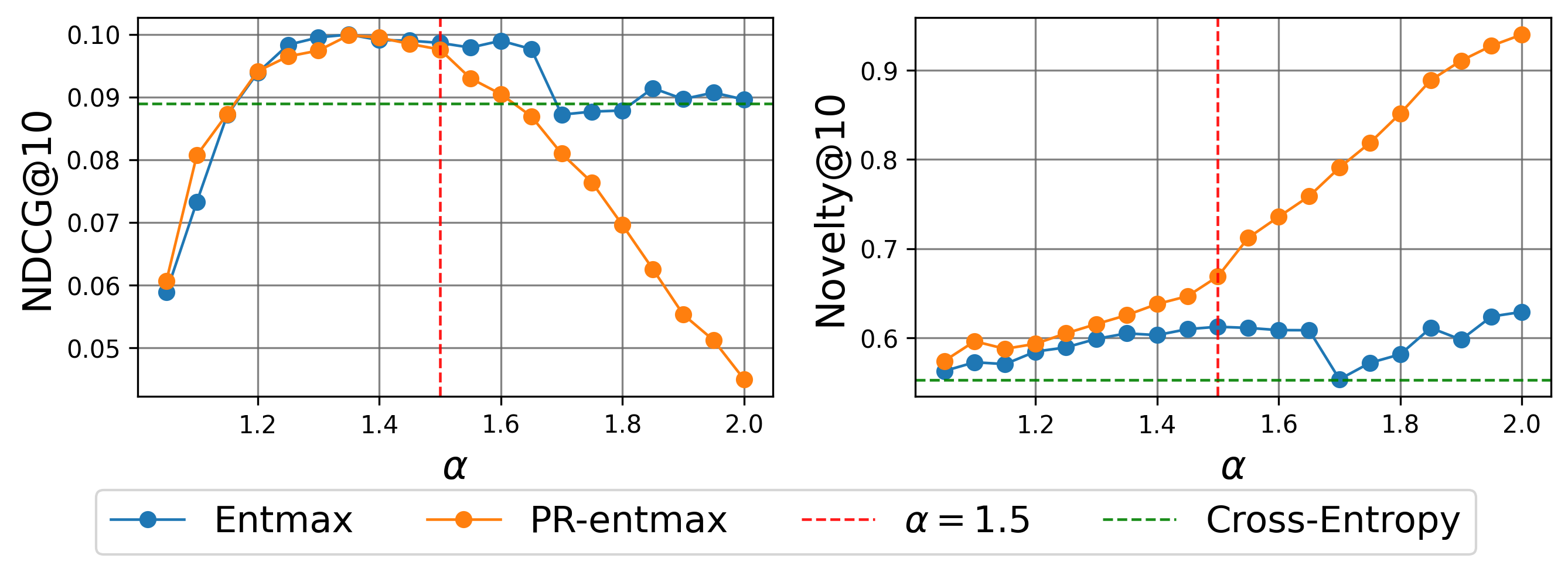}
    \caption{Yambda-50M}
    \end{subfigure}
    \begin{subfigure}{0.49\linewidth}
    \includegraphics[width=\textwidth]{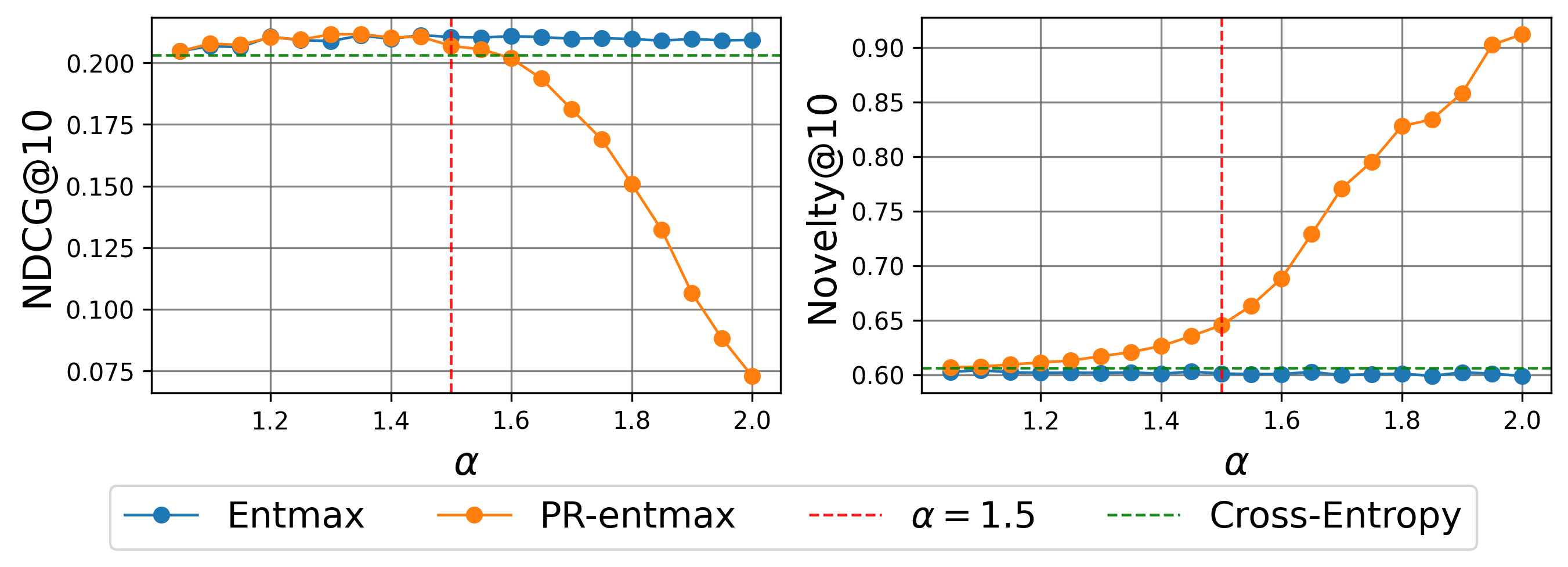}
    \caption{Yoochoose}
    \end{subfigure}
    \begin{subfigure}{0.49\linewidth}
    \includegraphics[width=\textwidth]{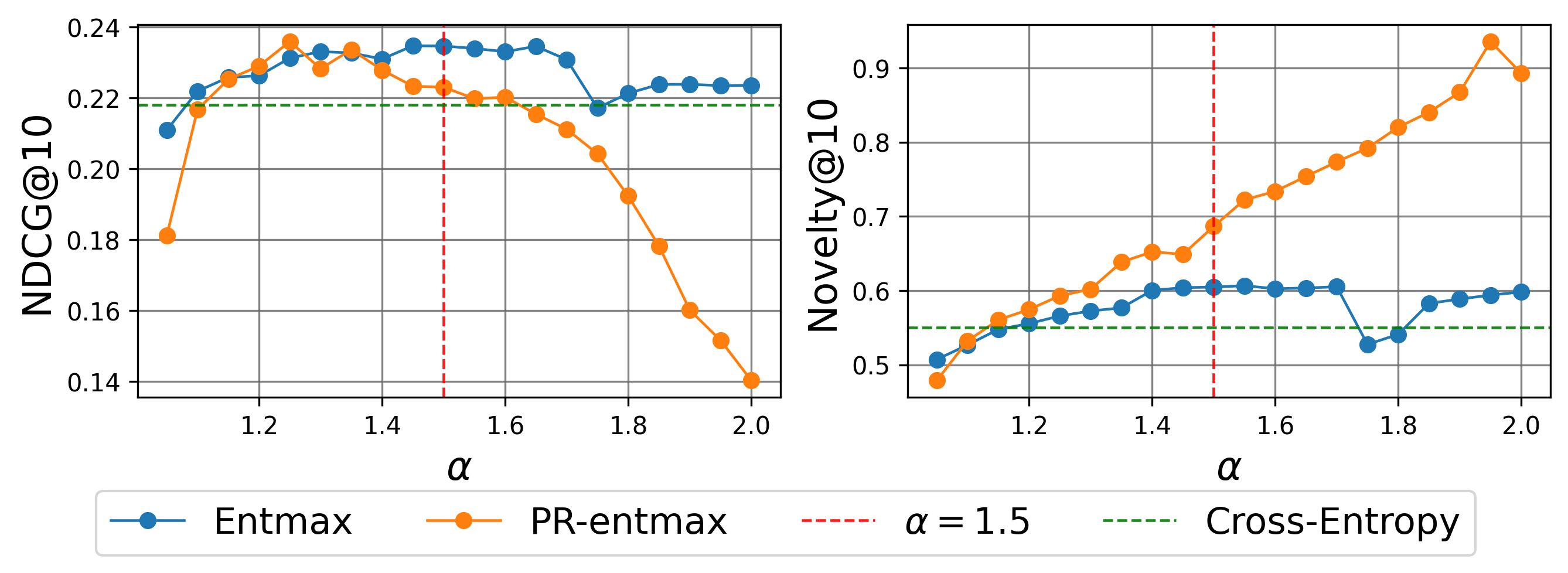}
    \caption{Zvuk}
    \end{subfigure}
    \caption{Impact of the sparsity parameter $\alpha$ on recommendation metrics NDCG@10 and Novelty@10 for Entmax and PR-entmax loss functions. Setting $\alpha=1.5$ achieves near‑optimal NDCG while maintaining high novelty.}
    \label{fig:sparsity}
\end{figure*}

Popularity debiasing methods often reduce bias at the cost of ranking quality. To analyze this relevance–debiasing trade-off, we evaluate each method across different values of the hyperparameter $\gamma$, which controls the strength of popularity correction, and collect the resulting operating points in the NDCG–Novelty space. This allows us to compare not only a single tuned configuration, but the full trade-off frontier induced by each approach. Figure~\ref{fig:pareto} illustrates the NDCG@10 versus Novelty@10 trade-off. The standard cross-entropy loss serves as the main reference point for comparison.

PR-entmax shifts the Pareto front upward, improving both metrics simultaneously on all datasets. On the Zvuk dataset, all other debiasing approaches exhibit relevance degradation as novelty increases. On other datasets, some baselines admit regimes where Novelty improves without sacrificing NDCG. In particular, such configurations are observed for Pop Sampling and Logit Adjustment on Tmall, for Pop Sampling, Logit Adjustment, and IPW on Yambda-50M, and for Pop Sampling on Yoochoose. Nevertheless, PR-entmax achieves a more favorable relevance-debiasing trade-off, simultaneously reaching higher Novelty and NDCG levels.

These results suggest that the proposed sparse prior-regularized objective improves not only individual operating points, but the overall Pareto frontier of relevance and popularity debiasing.

\subsection{Exposure Analysis}
\label{sec:exposure}

To further analyze the debiasing behavior of different objectives, we perform an exposure analysis based on Lorenz-style exposure curves. For each method, the x-axis represents the fraction of the most popular items in the catalog, while the y-axis shows the cumulative fraction of total recommendation exposure allocated to these items. Such curves provide a more detailed view of popularity bias than aggregate metrics such as Novelty or Coverage, since they characterize how recommendations are distributed across the popularity spectrum.

Figure~\ref{fig:exposure} presents the exposure curves for the standard cross-entropy baseline, PR-entmax, its limiting cases Entmax and Logit Adjustment, as well as Pop Sampling, which demonstrates one of the strongest relevance–debiasing trade-offs among the considered baselines. The remaining methods are omitted for clarity in order to avoid excessive visual clutter.

All considered debiasing approaches lie below the standard cross-entropy baseline, indicating reduced exposure concentration on the most popular items. Interestingly, Entmax without prior correction also improves the exposure of less popular items on three out of four datasets, although the effect remains substantially weaker than for the proposed approach.

For Yambda-50M and Zvuk, the exposure curve of PR-entmax lies clearly below the competing methods over most of the popularity range, indicating that the proposed objective allocates substantially more exposure to mid- and tail-popularity items. On the remaining datasets, the relative ordering between debiasing methods varies across different regions of the distribution. Nevertheless, PR-entmax consistently reduces exposure concentration on the most popular items while maintaining exposure levels for tail items comparable to the strongest competing approaches.

\subsection{Sensitivity to the Sparsity Parameter}
\label{sec:sparsity}

Tsallis entropy, and consequently the proposed variational objective, introduces the hyperparameter $\alpha$, which controls the sparsity of the resulting probability mapping. In all previous experiments, we used $\alpha = 1.5$. In this section, we analyze how different values of $\alpha$ affect recommendation relevance and popularity debiasing, and justify this choice empirically.

Figure~\ref{fig:sparsity} illustrates the dependence of NDCG@10 and Novelty@10 on the sparsity parameter for $\alpha$ values in the interval $[1.05, 2]$. In the limiting case $\alpha \to 1$, the mapping approaches softmax, while $\alpha = 2$ corresponds to sparsemax. The figure reports the results both for Entmax without prior correction and for the proposed PR-entmax objective. The standard Cross-Entropy baseline is shown as a dotted horizontal line.

Across most values of $\alpha$, Entmax remains above the cross-entropy baseline on both NDCG@10 and Novelty@10, except for several points near the low- and high-sparsity extremes. However, the resulting novelty improvements remain relatively modest. In contrast, PR-entmax achieves substantially stronger debiasing gains across the entire range of sparsity values.

For highly sparse solutions, Novelty@10 increases dramatically, but this improvement comes at the cost of a substantial reduction in ranking quality. At the same time, there exists a relatively broad interval of $\alpha$ values, approximately between $1.2$ and $1.6$, where PR-entmax simultaneously improves NDCG@10 while maintaining strong debiasing performance across all datasets. This behavior indicates that the proposed approach is stable and not overly sensitive to the precise choice of the sparsity parameter. The selected value $\alpha = 1.5$ achieves near-optimal ranking quality while preserving substantial debiasing improvements.

\section{Conclusion}

In this work, we revisited popularity bias in sequential recommendations from the perspective of the training objective. We showed that making popularity correction context-dependent through a sparse probability mapping provides an effective and principled alternative to standard softmax-based training.
Our experiments further show that sparsity alone, through Entmax, consistently improves ranking accuracy but has a mixed, dataset-dependent effect on popularity bias. By combining sparse competition with popularity-aware prior correction, PR-entmax improves both recommendation relevance and popularity debiasing, yielding a more favorable relevance–debiasing trade-off.
The proposed approach operates purely at the level of the training objective and does not require architectural modifications, making it applicable to a wide range of recommendation models. Future work may further explore sparse transformations for output distributions, extend the framework to other recommendation settings, and investigate different forms of prior correction.

\appendix
\section{Derivation of the probability mapping}
\label{sec:appendix}

To solve the variational problem (8) and obtain an explicit
expression for the probability mapping $\mathbf{p}^*(\mathbf{z})$,
we introduce Lagrange multipliers for the equality constraint
$\sum_i p_i = 1$ and the inequality constraints $p_i \geq 0$, and
apply the Karush--Kuhn--Tucker (KKT) conditions. The corresponding
Lagrangian is
\begin{align*}
\mathcal{L}(\mathbf{p},\lambda,\boldsymbol{\mu})
&=
\sum_i p_i \bigl(z_i + \gamma \log q_i\bigr)
+
\frac{\theta}{\alpha(\alpha-1)}
\left(1-\sum_i p_i^\alpha\right)
\nonumber\\
&\quad
+\lambda\left(1-\sum_i p_i\right)
+\sum_i \mu_i p_i .
\end{align*}

The optimal solution must satisfy the KKT conditions
\begin{equation*}
p_i \geq 0,
\qquad
\mu_i \geq 0,
\qquad
\mu_i p_i = 0,
\qquad
\sum_i p_i = 1.
\end{equation*}
The stationarity condition with respect to $p_i$ gives
\begin{equation*}
0
=
\frac{\partial \mathcal{L}}{\partial p_i}
=
z_i + \gamma \log q_i
-\frac{\theta}{\alpha-1}p_i^{\alpha-1}
-\lambda+\mu_i .
\end{equation*}

The complementary slackness condition $\mu_i p_i = 0$ separates
the solution into active and inactive coordinates. For $p_i>0$,
we have $\mu_i=0$, and therefore
\begin{equation*}
p_i
=
\left[
\frac{\alpha-1}{\theta}
\left(z_i+\gamma\log q_i-\lambda\right)
\right]^{\frac{1}{\alpha-1}}.
\end{equation*}
For $p_i=0$, stationarity implies
\begin{equation*}
\mu_i
=
\lambda-\left(z_i+\gamma\log q_i\right)
\geq 0,
\end{equation*}
and hence
\begin{equation*}
z_i+\gamma\log q_i \leq \lambda.
\end{equation*}

Combining the active and inactive cases yields
\begin{equation*}
p_i^*(\mathbf{z})
=
\left[
\frac{\alpha-1}{\theta}
\left(z_i+\gamma\log q_i-\lambda\right)
\right]_+^{\frac{1}{\alpha-1}},
\end{equation*}
where $[x]_+ = \max(0,x)$. Finally, defining
$\tau=\lambda/\theta$, we obtain
\begin{equation*}
p_i^*(\mathbf{z})
=
\left[
(\alpha-1)
\left(
\frac{z_i+\gamma\log q_i}{\theta}-\tau
\right)
\right]_+^{\frac{1}{\alpha-1}},
\end{equation*}
where $\tau$ is chosen such that
$\sum_i p_i^*(\mathbf{z})=1$.

\section*{GenAI Usage Disclosure}
Generative AI tools were used during the preparation of this manuscript. Specifically, OpenAI ChatGPT and Anthropic Claude were used to assist with English wording, LaTeX editing, and table and figure formatting. No generative AI tool was used to produce experimental data, conduct experiments, compute reported metrics, or independently formulate the scientific claims or conclusions of the paper. All AI-assisted text and formatting suggestions were manually reviewed, edited, and verified by the authors. The authors remain fully responsible for the accuracy, originality, and integrity of the submitted work.

\bibliographystyle{ACM-Reference-Format}
\balance
\bibliography{sections/bibliography}

\end{document}